\documentclass[10pt,a4paper]{article} 
\usepackage{jcappub, natbib}

\allowdisplaybreaks

\usepackage{ifthen}
\usepackage{graphicx}% Include figure files
\usepackage{dcolumn}% Align table columns on decimal point
\usepackage{bm}% bold math
\usepackage{color}
\usepackage{amsmath}
\usepackage{amssymb}
\usepackage{subfigure}

\newcommand{\vp}{\mathbf{p}}
\newcommand{\vq}{\mathbf{q}}

\newcommand{\vx}{\mathbf{x}}
\newcommand{\vy}{\mathbf{y}}
\newcommand{\vk}{\mathbf{k}}

\newcommand{\himpc}{{\hbox {$~h^{-1}$}{\rm ~Mpc}}}
\newcommand{\hmpci}{{\hbox {$~h{\rm ~Mpc}^{-1}$}}}
\newcommand{\be}{\begin{equation}}
\newcommand{\ee}{\end{equation}}

\begin{document}

\title{Distribution function approach to redshift space distortions. Part II: $N$-body simulations}

\author[a]{Teppei Okumura,} \emailAdd{teppei@ewha.ac.kr}
\author[a,b,c,d]{Uro{\v s} Seljak,} \emailAdd{useljak@berkeley.edu}
\author[b,e]{Patrick McDonald,} \emailAdd{pvmcdonald@lbl.gov}
\author[d]{and Vincent Desjacques} \emailAdd{dvince@physik.uzh.ch}

\affiliation[a]{Institute for the Early Universe, Ewha Womans
  University, Seoul 120-750, S. Korea}

\affiliation[b]{Department of Physics and Lawrence Berkeley National
  Laboratory, University of California, Berkeley, California 94720,
  USA} 

\affiliation[c]{Department of Astronomy, University of California,
  Berkeley, California 94720, USA} 
  
\affiliation[d]{Institute of Theoretical Physics, University of Zurich, 8057 Zurich, Switzerland}

\affiliation[e]{Physics Dept., Brookhaven National Laboratory,
  Building 510A, Upton, NY 11973-5000, USA}

\abstract{ Measurement of redshift-space distortions (RSD) offers an
  attractive method to directly probe the cosmic growth history of
  density perturbations.  A distribution function approach where RSD
  can be written as a sum over density weighted velocity moment
  correlators has recently been developed.  In this paper we use
  results of N-body simulations to investigate the individual
  contributions and convergence of this expansion for dark matter.  If
  the series is expanded as a function of powers of $\mu$, cosine of
  the angle between the Fourier mode and line of sight, then there are
  a finite number of terms contributing at each order. We present
  these terms and investigate their contribution to the total as a
  function of wavevector $k$. For $\mu^2$ the correlation between
  density and momentum dominates on large scales. Higher order
  corrections, which act as a Finger-of-God (FoG) term, contribute 1\%
  at $k \sim 0.015\hmpci$, 10\% at $k \sim 0.05\hmpci$ at $z=0$, while
  for $k>0.15\hmpci$ they dominate and make the total negative. These
  higher order terms are dominated by density-energy density
  correlations which contributes negatively to the power, while the
  contribution from vorticity part of momentum density
  auto-correlation adds to the total power, but is an order of
  magnitude lower. For $\mu^4$ term the dominant term on large scales
  is the scalar part of momentum density auto-correlation, while
  higher order terms dominate for $k>0.15\hmpci$. For $\mu^6$ and
  $\mu^8$ we find it has very little power for $k<0.15\hmpci$,
  shooting up by 2-3 orders of magnitude between $k<0.15\hmpci$ and
  $k<0.4\hmpci$.  We also compare the expansion to the full 2-d
  $P^{ss}(k,\mu)$, as well as to the monopole, quadrupole, and
  hexadecapole integrals of $P^{ss}(k,\mu)$.  For these statistics an
  infinite number of terms contribute and we find that the expansion
  achieves percent level accuracy for $k\mu<0.15 \hmpci$ at 6-th
  order, but breaks down on smaller scales because the series is no
  longer perturbative.  We explore resummation of the terms into FoG
  kernels, which extend the convergence up to a factor of 2 in scale.
  We find that the FoG kernels are approximately Lorentzian with
  velocity dispersions around $600$km/s at $z=0$.  } 
%\pacs{}
\arxivnumber{1109.1609}

\maketitle

%%%%%%%%%%%%%%%%%%%%%%%%%%%%%%%%%%%%%%%%%%%%%%
% Section 1 Introduction
%%%%%%%%%%%%%%%%%%%%%%%%%%%%%%%%%%%%%%%%%%%%%%
\section{Introduction}\label{sec:intro}

Galaxy redshift surveys are one of the most powerful tools to probe
cosmological models \cite{Peebles:1980}. One way to extract the
information is from the shape of the power spectrum or correlation
function, assuming it traces the underlying dark matter. Another
method involves baryonic acoustic oscillations (BAOs), detected in
various redshift surveys, which enables one to measure angular
diameter distance and compare it to the same quantity measured in
cosmic microwave background. This in turn probes the expansion history
of the Universe and allows to study the nature of dark energy
\cite[e.g.,][]{Eisenstein:2005, Cole:2005}.  Third piece of
information in redshift surveys comes from redshift space distortions
(RSD): the observed galaxy distribution is distorted along the line of
sight due to the Doppler shifts caused by peculiar velocities
\cite{Jackson:1972, Kaiser:1987, Hamilton:1998}. In linear theory this
allows one to measure the rate of growth of structure, which allows
for another way to measure the matter content of the universe,
including the amount and nature of dark energy.  The last two methods
are complementary: cosmological models in different gravity theories
with the same expansion history cannot be distinguished by the
distance scales of BAOs, but can if growth of structure is also
measured \cite[e.g.,][]{Linder:2005, Jain:2008, Song:2009}.
Additional information is obtained by Alcock-Paczy\'{n}ski test
\cite{Alcock:1979, Matsubara:1996, Ballinger:1996}.

RSD have been analyzed in many galaxy surveys to determine the
cosmological models \citep[e.g.,][]{Peacock:2001, Zehavi:2002,
  Hawkins:2003, Tegmark:2004, Tegmark:2006, Ross:2007, Guzzo:2008,
  Okumura:2008, Cabre:2009, Blake:2011}.  However, it was shown by
\cite{Tinker:2006, Okumura:2011, Jennings:2011, Kwan:2011} that the
growth rate reconstructed from the redshift-space distortions can have
scale dependent biases, which indicate a breakdown of linear theory
predictions. These effects show up on relatively large scales,
suggesting one must go beyond the linear theory in the analysis of
RSD.  This will become even more important in the future, with several
ongoing and upcoming galaxy surveys that will measure RSD to a high
precision \cite{Hill:2004, Glazebrook:2007, Schlegel:2009,
  Sumiyoshi:2009, Schlegel:2009a}.

Given the high precision of the future surveys, correspondingly more
accurate theoretical predictions become essential for their
interpretation.  As was emphasized by \cite{Scoccimarro:2004}, there
are important nonlinear effects that need to be addressed in order to
achieve accurate theoretical predictions.  In order to account for the
nonlinearity of the gravitational evolution, standard perturbation
theory has long been used to describe the power spectrum at
quasi-nonlinear scales \cite[e.g.,][]{Bernardeau:2002}.  Recently
there have been many studies to predict the power spectrum in
nonlinear regime beyond the framework of the standard perturbation
theory (SPT) \cite{Crocce:2006, Crocce:2006b, Matarrese:2007,
  McDonald:2007, Valageas:2007, Taruya:2008}.  These different
approaches were compared to the $N$-body simulations in
\cite{Nishimichi:2009, Carlson:2009}.  Similarly, initial RSD work was
based on the lowest order SPT \cite{Heavens:1998, Scoccimarro:1999,
  Bharadwaj:2001, Pandey:2005}.  However, as pointed out by
\cite{Scoccimarro:1999, Scoccimarro:2004}, SPT in redshift space
breaks down at larger scales than in real space because of nonlinear
redshift distortion effects, sometimes called Finger-of-God (FoG)
effect \cite{Jackson:1972}.  Recent development using more
sophisticated perturbation methods applicable to the redshift-space
power spectrum includes \cite{Matsubara:2008, Taruya:2010,
  Valageas:2011}.  \cite{Blake:2011} present detailed comparisons of
these predictions with the observed galaxy data.

Recent paper \cite{Seljak:2011} has shown that one can write the
Fourier mode of density in redshift space as a sum over mass weighted
moments of radial velocity, which are integrals of powers of velocity
over the momentum part of the phase space distribution function. The
corresponding RSD power spectrum can be written as a sum over auto and
cross-correlators of these moments. This series always converges for
sufficiently small expansion parameter defined below. We will use the
Fourier description in this paper and scale is expressed in terms of
wavevector amplitude $k$, while angular dependence is expressed in
terms of $\mu$, cosine of the angle between the line of sight and
Fourier mode.  The expansion parameter depends on the product of the
two $k_{\parallel}=k\mu$.  It has been shown in \cite{Seljak:2011}
that the moments can be decomposed into helicity eigenstates, which
are eigenmodes under rotation around direction of $\vk$ vector. Only
equal helicity eigenstates correlate, leading to a specific angular
structure of the correlators. This analysis shows that if one expands
the series into powers of $\mu^2$, a finite number of terms contribute
at each (finite) order. This suggests that RSD can be better
understood in terms of this expansion rather than the Legendre moments
usually used \cite{Seljak:2011}.  On the other hand, Legendre moments
are uncorrelated in real observations, while powers of $\mu^2$ are
not, leading to correlations between the higher and lower orders.  We
will pursue both approaches here.

This is the second paper in a series studying the redshift-space
distortions based on a distribution function approach, following the
theory and angular decomposition presented in \cite{Seljak:2011}.  In
this paper we test the formalism to describe the redshift-space power
spectrum in nonlinear regime using a large set of cosmological
$N$-body simulations, as well as present the individual terms of
expansion for comparison against each other.  We focus on the dark
matter in this paper, leaving the application to halos and galaxies to
future work.  The structure of this paper is as follows.  In section
\ref{sec:theory} we briefly describe the distribution function
approach to RSD.  Then we apply it to simulations to test this
expansion and show the contributions from individual terms: in section
\ref{sec:analysis} we first show the contributions to the 2-d power
spectrum in redshift space, then proceed to Legendre moments.  We
discuss the FoG modeling in section \ref{sec:fog_resum} and present an
attempt to compare the expansion to one in terms of volume weighted
quantities.  Finally in section \ref{sec:angular} we apply the method
to powers of $\mu^2$ expansion, which we argue is a more natural way
to expand 2-d information, showing individual contributions to the
lowest order terms.  This is followed by conclusions in section
\ref{sec:conclusion}.

%%%%%%%%%%%%%%%%%%%%%%%%%%%%%%%%%%%%%%%%%%%%%%
% Section 2 RSD
%%%%%%%%%%%%%%%%%%%%%%%%%%%%%%%%%%%%%%%%%%%%%%

\section{Redshift-space distortions from the distribution function}
\label{sec:theory}
The exact evolution of collisionless particles is described by the
Vlasov equation \cite{Peebles:1980}.  Following the discussion by
\cite{McDonald:2011}, we start from the distribution function of
particles $f(\vx,\vq,t)$ at phase-space position $(\vx,\vq)$ in order
to derive the perturbative redshift-space distortions.  Here $\vx$ is
the comoving position and $\vq=\vp/a$ is the comoving momentum ($\vp$
is the proper momentum and $a$ is the scale factor). 
The density field in redshift space is related to moments of distribution function as
\begin{equation}
  \delta_s(\vk)=
  \sum_{L=0}\frac{1}{L!}
  \left(\frac{i k_\parallel}{H}\right)^L T_\parallel^L(\vk) ~,
  \label{eq:deltak}
\end{equation}
where $H$ is the Hubble parameter and $T_\parallel^L(\vk)$ is the
Fourier transform of $T_\parallel^L(\vx)$, defined as \be
T_\parallel^L(\vx)={m \over \bar{\rho}} ~ \int d^3\vq~
f\left(\vx,\vq\right) u_\parallel^L= \left\langle
\left(1+\delta(\vx)\right) u_\parallel^L(\vx)
\right\rangle_{\vx}, \label{eq:q_def} \ee where $u_\parallel$ is the
radial comoving velocity, $mu_\parallel=q_\parallel = \vq\cdot
\hat{r}$, $m$ is the particle mass, $\hat{r}$ is the unit vector
pointing along the observer's line of sight and $\bar{\rho}$ is the
mean mass density.

The power spectrum in redshift space is given by \cite{Seljak:2011}
\begin{equation}
  P^{ss}(\vk)=\sum_{L=0}^{\infty}\frac{1}{L!^2}\left(\frac{
    k\mu}{H}\right)^{2L} P_{LL}(\vk) +
  2\sum_{L=0}^{\infty}\sum_{L'>L}\frac{\left(-1\right)^{L}}{L!~L'!}
  \left(\frac{i k\mu}{H}\right)^{L+L'} P_{LL'}(\vk) \label{eq:p_ss} ~,
\end{equation}
where $k_{||}/k=\cos \theta=\mu$. 
It is useful to compare this to Kaiser's linear theory prediction
\cite{Kaiser:1987, Scoccimarro:2004}.  Thus we have
\begin{eqnarray}
  P^{ss}_{\rm Kaiser}(\vk)&=& \left\{
  \begin{array}{ll}
    \left( 1+f\mu^2 \right)^2 P_{\rm lin}(k)  & {\rm ; linear}, \\
    P_{00}+2f\mu^2 \left(\frac{ik}{H\mu f}\right)P_{01} + f^2\mu^4 \left( \frac{k}{H\mu f}\right)^2 P_{11} & {\rm ; nonlinear}, 
  \end{array}\label{eq:nl_kaiser}  
  \right.
\end{eqnarray}
where $P_{\rm lin}$ denotes the linear power spectrum and 
$f=d\ln{D}/d\ln{a}$ with $D$ the growth factor. 
These terms will in general have nonlinear corrections, so
we call this approximation the nonlinear Kaiser order approximation.
Replacing these lowest 3 moments with the standard linear theory we
obtain the original linear Kaiser model of equation
(\ref{eq:nl_kaiser}).  Here we want to view this series simply as a
series in $k_\parallel$, investigating the convergence as more terms
are added.

Note that the calculations never require anything but simple power
spectra of mass-weighted powers of velocity to be computed from the
simulations.  As we will compare RSD power spectrum to the sum from
individual terms there should not be much sampling variance in the
comparison, because both are calculated from the same simulation, so
the large-scale fluctuations will be the same. The order of
$k_\parallel=k\mu$ needed for convergence to a given level of accuracy
will inevitably increase as one goes to increasingly small scales,
with the whole expansion eventually breaking down once $k\mu
\sigma/H>1$, where $\sigma$ is a typical (comoving) velocity of the
system. We will see in section \ref{sec:analysis} that the nonlinear terms of $P_{LL'}$, 
particularly $P_{02}$, have significant contributions even at the scales 
larger than the breakdown scale. 

\subsection{Angular dependence} \label{sec:th_angular}

By performing helicity decomposition \cite{Seljak:2011} show that the
power spectrum can be written as 
\be
P_{LL'}(\vk)=\sum_{(l=L,L-2,..)}\sum_{(l'=L',L'-2,..;\; l'\ge
  l)}\sum_{m=0}^{l}P^{L,L',m}_{l,l'}(k)P_l^m(\mu)P_{l'}^m(\mu) ,
\label{pll}
\ee 
where $P_l^m(\mu)$ are the associated Legendre polynomials, which
determine the angular dependence of the spherical harmonics. There are
5 numbers that describe these objects: $L$ and $L'$ describe the power
of two velocity moments we are correlating, $l$, $l'$ describe the
rank of the object, for example $l=1$ is rank-1, which is a 3-d
vector, $l=2$ is a 3-d tensor etc.  Finally, $m$ is the helicity
eigennumber, which ranges between 0 and $l$ ($l \le l'$). Only equal
helicity components of expansion have a non-vanishing correlator.
There is a close relation between the order of the moments and their
angular dependence.  The lowest contribution in powers of $\mu$ to
$P^{ss}(k)$ is $\mu^{L+L'}$ if $L+L'$ is even or $\mu^{L+L'+1}$ if
$L+L'$ is odd, and the highest is $\mu^{2(L+L')}$. Thus for
$P_{00}(\vk)$ the only angular term is isotropic term ($\mu^0$), for
$P_{01}(\vk)$ the only angular term is $\mu^2$, $P_{11}(\vk)$ and
$P_{02}(\vk)$ contain both $\mu^2$ and $\mu^4$ etc.  Note that only
even powers of $\mu$ enter in the final expression, as required by the
symmetry.  We can thus write
\begin{equation}
  P^{ss}(\vk)=\sum_{L=0}^{\infty}\frac{1}{L!^2}\left(\frac{
    k}{H}\right)^{2L} \sum_{j=2L}^{4L}P^{(j)}_{LL}(\vk)\mu^{j} +
  2\sum_{L=0}^{\infty}\sum_{L'>L}\frac{\left(-1\right)^{L}}{L!~L'!}
  \left(\frac{i k}{H}\right)^{L+L'} \sum_{j=(L+L') {\rm or}
    (L+L'+1)}^{2(L+L')}P^{(j)}_{LL'}(\vk)\mu^{j} \label{eq:p_ss_angular} ~,
\end{equation}
so that terms $P^{(j)}_{LL'}$ are coefficients in expansion in powers
of $\mu^j$ of contributions of $L,L'$ terms to $P^{ss}$. Since the $j$
index has to be even, the lowest order is either $L+L'$ or
$L+L'+1$, whichever is even, thus the higher order terms also have 
higher order powers of $\mu^j$ and one can in principle separate them 
from lower order terms. These terms can be uniquely extracted
from simulations from angular dependence of $P_{LL'}$ terms and so we
will focus on them, although sometimes it is useful to decompose them
into the individual helicity eigenstates instead.

%%%%%%%%%%%%%%%%%%%%%%%%%%%%%%%%%%%%%%%%%%%%%%
% Section 3 Numerical Analysis
%%%%%%%%%%%%%%%%%%%%%%%%%%%%%%%%%%%%%%%%%%%%%%
\section{Numerical analysis}\label{sec:analysis}

\subsection{$N$-body simulations}\label{sec:nbody}

The power spectra of the derivative expansion are all from
mass-weighted velocity moments and thus can be straightforwardly
measured from simulations.  We use a series of $N$-body simulations of
the $\Lambda$CDM cosmology seeded with Gaussian initial conditions,
which is an updated version of \citep{Desjacques:2009}.  The
primordial density field is generated using the matter transfer
function by CMBFAST \cite{Seljak:1996}.  We adopt the standard
$\Lambda$CDM model with the mass density parameter $\Omega_m=0.279$,
the baryon density parameter $\Omega_b=0.0462$, the Hubble constant
$h=0.7$, the spectral index $n_s=0.96$, and a normalization of the curvature 
perturvations $\Delta_{\cal R}^2=2.21\times 10^{-9}$ (at $k=0.02~{\rm Mpc}^{-1}$)
which gives the density fluctuation
amplitude $\sigma_8\approx 0.81$, which are the best-fit parameters in the
WMAP 5-year data \cite{Komatsu:2009}. We employ $1024^3$ particles of
mass $3.0\times 10^{11} h^{-1}M_\odot$ in a cubic box of side
$1600\himpc$.  The positions and velocities of all the dark matter
particles are output at $z=0,~0.509,~0.989$, and 2.070, which are
quoted as $z=0,~0.5,~1$, and 2 in what follows for simplicity.  We use
12 independent realizations in order to reduce the statistical
scatters.  For the detail of the simulations see
\cite{Desjacques:2009}.

Next we describe how we measure the power spectra from our simulation
samples.  We assign the density field and the mass-weighted velocity
moments in real space on $1024^3$ grids using a cloud-in-cell
interpolation method according to the positions of particles.  To
directly measure $P^{ss}(\vk)$ we also need the density field in
redshift space.  In measuring the redshift-space density field, we
distort the positions of particles along the line-of-sight according
to their peculiar velocities before we assign them to the grid.  We
regard each direction along the three axes of simulation boxes as the
line of sight and the statistics are averaged over three projections
of all realizations for a total of 36 samples.  We use a fast Fourier
transform to measure the Fourier modes of the density fields in real
space $\delta(\vk)$ and in redshift space $\delta_s(\vk)$, as well as
the mass-weighted velocity moment fields in real space
$T_\parallel^L(\vk)$.  Finally, the power spectrum in redshift space,
$P^{ss}(\vk)$, as well as the power spectra of the mass-weighted
velocity moments $P_{LL'}(\vk)$, are measured by multiplying the modes
of the two fields (or squaring in case of auto-correlation) and
averaging over the Fourier modes.  Throughout this paper, we neglect
shot noise because we have sufficient number of dark matter particles
and such an effect is thus negligibly small.  Error bars in the
following results are estimated from bootstrap resampling.  The
dispersion in power spectra measurements is large on large scales
because of sampling variance, but it is mostly eliminated by taking
the ratio of the two spectra obtained from the same set of
realizations (e.g. \cite{McDonald:2009}).

\subsection{Matter power spectrum}\label{sec:power_measure}

\begin{figure}[!t]
\includegraphics[width=1.0\textwidth]{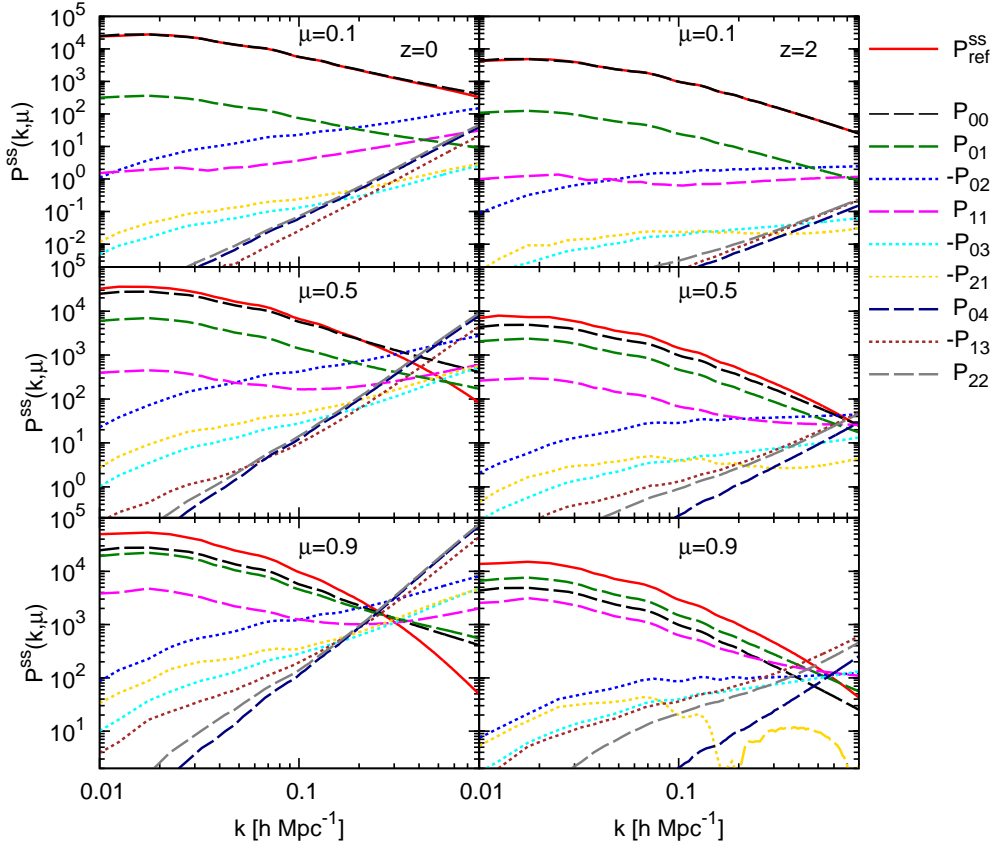}
\caption{Power spectra measured in redshift space $P^{ss}(k,\mu)$ and
  individual contributions to it from the terms of the moments
  expansion up to 4-th order at $z=0$ (left) and $z=2$ (right).  The
  width of $\mu$ bin is 0.2 centered around the values shown in each
  panel.  The solid and dashed lines show the positive values, while
  the dotted lines the negative values.  }
\label{fig:pkmu_individual}
\end{figure}

In this subsection we first measure the redshift-space power spectrum,
$P^{ss}(k,\mu)$, directly in redshift space, which we call the
``reference" power spectrum.  The reference power spectrum in redshift
space is shown as functions of $(k, \mu)$ at $z=0$ and 2 as the red
lines in figure \ref{fig:pkmu_individual}.  We adopt the constant
$\mu$ binning into five bins between $0\leq \mu\leq 1$, but only three
$\mu$ bins among the five are plotted.  In figure
\ref{fig:pkmu_individual} we also show contributions of the terms of
$P_{LL'}$ for $(0\leq L+L'\leq 4)$ to $P^{ss}(k,\mu)$ computed from
the mass-weighted velocity moments.  At $\mu\sim 0$ contributions from
the higher order power spectra of the velocity moments are small and
$P^{ss}\simeq P_{00}$ because each $P_{LL'}$ is multiplied by a factor
of $(k\mu)^{L+L'}$.

%%%%%%%%%%%%%%%%%%%%%%%%%%%%%%%%%%%%%%%%%%%%%%
% Figure 1
%%%%%%%%%%%%%%%%%%%%%%%%%%%%%%%%%%%%%%%%%%%%%%
%%%%%%%%%%%%%%%%%%%%%%%%%%%%%%%%%%%%%%%%%%%%%%

On large scales one expects $P_{00}$ to be followed by the other two
linear order terms, which are $P_{01}$ and the scalar part of
$P_{11}$, i.e. $P_{11}^{110}$. Note however that the latter scales as
$\mu^4$, while there are two nonlinear terms, the vector part of
$P_{11}$, $P_{11}^{111}$, and the scalar part of $P_{02}$, which
itself has two terms, both helicity 0, one from energy density
correlated with the density $P_{00}^{020}$, and one from anisotropic
stress density correlated with density, $P_{02}^{020}$, that contain
terms which scale as $\mu^2$. As a result, {\it for sufficiently low
  $\mu$ these nonlinear terms dominate over the linear term in
  $P_{11}^{110}$ even on very large scales}. For example, for
$\mu=0.1$ at $z=0$ we see that $P_{02}$ dominates over $P_{11}$ on all
scales probed, despite the fact that $P_{11}$ contains a linear order
term. We also see that $P_{11}$ does not follow the linear theory on
all but the largest scales, but instead has the shape dependence
similar to $P_{02}$, characteristic of the nonlinear terms. As pointed
out in \cite{Seljak:2011}, the nonlinear helicity 1 (vector or
vorticity) terms in $P_{11}$ are closely connected to $P_{02}$ and
partially cancel each other.  This angular decomposition is discussed
further below in section \ref{sec:angular}, where we present the
individual helicity terms separately. 

Because of $(k\mu)^{L+L'}$ weight the higher-order terms scale more
rapidly with $k$, and dominate on small scales: this is the region
where RSD are dominated by FoG effects.  One needs to take into
account more and more higher-order terms in order to make the
expansion (equation (\ref{eq:p_ss})) valid at such smaller scales.  This
effect is more significant at $z=0$ due to higher velocities. One can
see that the higher order terms cross the lower order terms at $k\mu
\sim 0.2\hmpci$ ($z=0$) and $k \mu \sim 0.4\hmpci$ ($z=2$). This is
where the perturbative parameter $k\mu \sigma/H$ becomes of order
unity and the perturbative approach breaks down,  consistent with  \cite{Scoccimarro:2004}. At that point higher
order terms dominate over the lower order terms and we no longer have
a convergence. This can be seen in figure \ref{fig:pkmu_individual}:
while for high $k$ the redshift space power spectrum $P^{ss}(k,\mu)$
decreases in power relative to the real space case $P_{00}(k)$, the
individual terms in the series expansion increase due to their
$(k\mu)^{L+L'}$. This suggests that a non-perturbative approach is
needed in this regime: we will explore the so-called FoG resummation
in section \ref{sec:fog_resum}.  Figure \ref{fig:pkmu_individual}
suggests that the typical velocity $\sigma$ in the expansion is about
500km/s at $z=0$ and 250km/s at $z=2$. We confirm these numbers in a
more detailed FoG analysis below.

\subsection{Legendre expansion}\label{sec:legendre}

We can compare the agreement between moments expansion and the full
$P^{ss}$ as a function of the order in the series $(L,L')$.  It is
customary to expand the redshift-space power spectrum in terms of
Legendre multipole moments \cite[e.g.,][]{Yamamoto:2005, Okumura:2011,
  Tocchini-Valentini:2011, Taruya:2011, Blake:2011}.  The motivation for this
expansion is that if one uses full angular information the Legendre
moments are uncorrelated.  Using Legendre polynomials ${\cal
  P}_l(\mu)$, we have \cite{Cole:1994}
\begin{equation}
  P^{ss}(k,\mu)=\sum_{l=0,2,4,\cdots}P^{ss}_l(k){\cal P}_l(\mu) ~.
\end{equation}
The multipole moments, $P^{ss}_l$, are obtained by inversion of this relation, 
\begin{equation}
  P^{ss}_l(k)=\frac{2l+1}{2}\int^{+1}_{-1}P^{ss}(k,\mu){\cal P}_l(\mu)d\mu ~. 
\end{equation}
Most of the studies on modeling redshift-space distortions
focus on the monopole ($l=0$) and quadrupole ($l=2$), although 
hexadecapole ($l=4$) also contains some cosmological information \cite{Taruya:2011}. 

\begin{figure}%[!t]
\centering
\subfigure{\includegraphics[width=1.\textwidth]{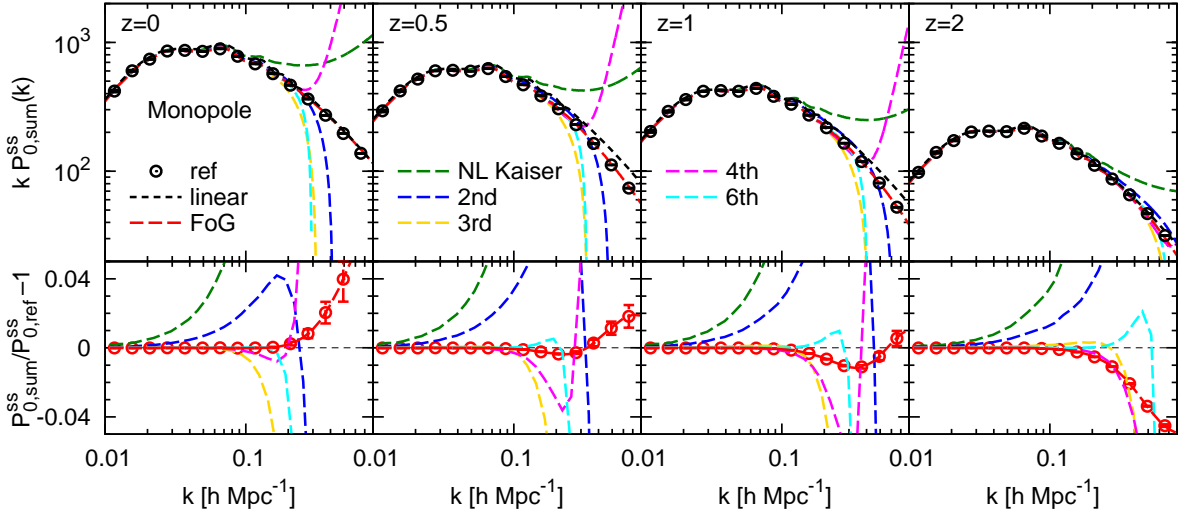}}
\caption{{\it Upper panels:} we show monopole moments of 
  power spectrum in redshfit space
  $P^{ss}$.  The vertical axis is multiplied by $k$ to
  clearly illustrate departures from a reference power spectrum.  The
  green, blue, yellow, magenta and cyan lines
  respectively show our model prediction up to nonlinear Kaiser, 2nd,
  3rd, 4th and 6th order corrections, measured from the
  simulations.  The black lines are linear theory prediction. 
  The black points with errorbars show the reference power spectrum. 
  The red lines show our FoG model (section \ref{sec:fog_resum}). 
  {\it Lower panels:} error
  between the summed power spectrum and the reference spectrum. The
  meaning of the color of each line is the same as that of the upper
  panels. For reference errorbars are shown for the result of our FoG model.   
}
\label{fig:pkl_mono}
\end{figure}

\begin{figure}%[!t]
\centering
\subfigure{\includegraphics[width=.49\textwidth]{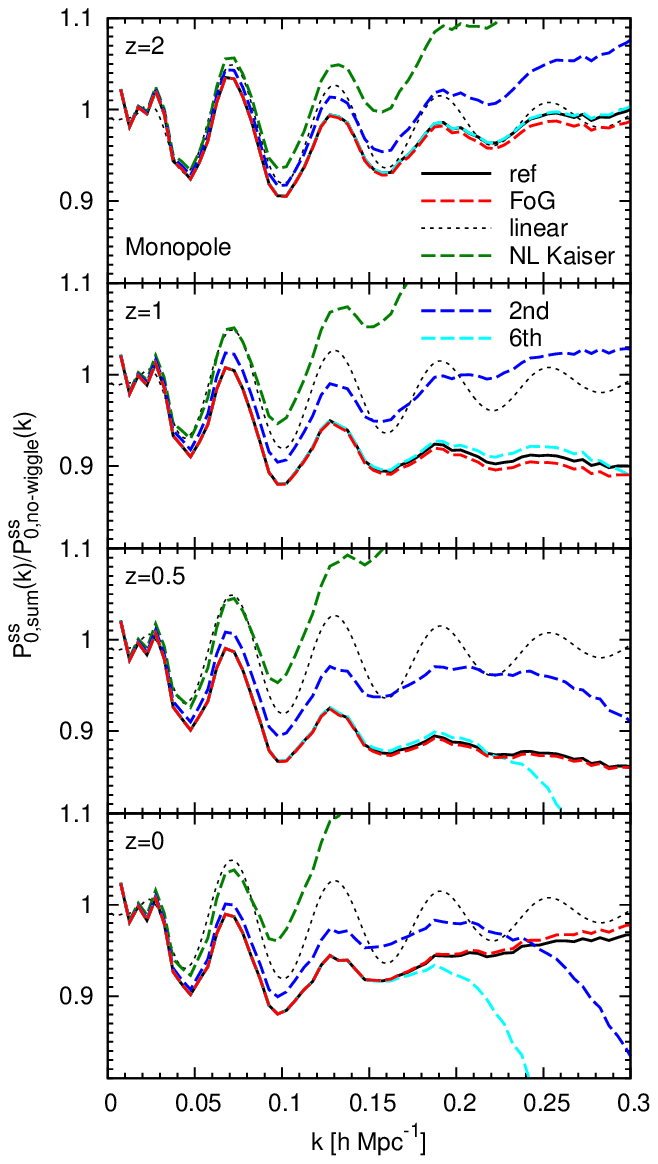}}
\subfigure{\includegraphics[width=.49\textwidth]{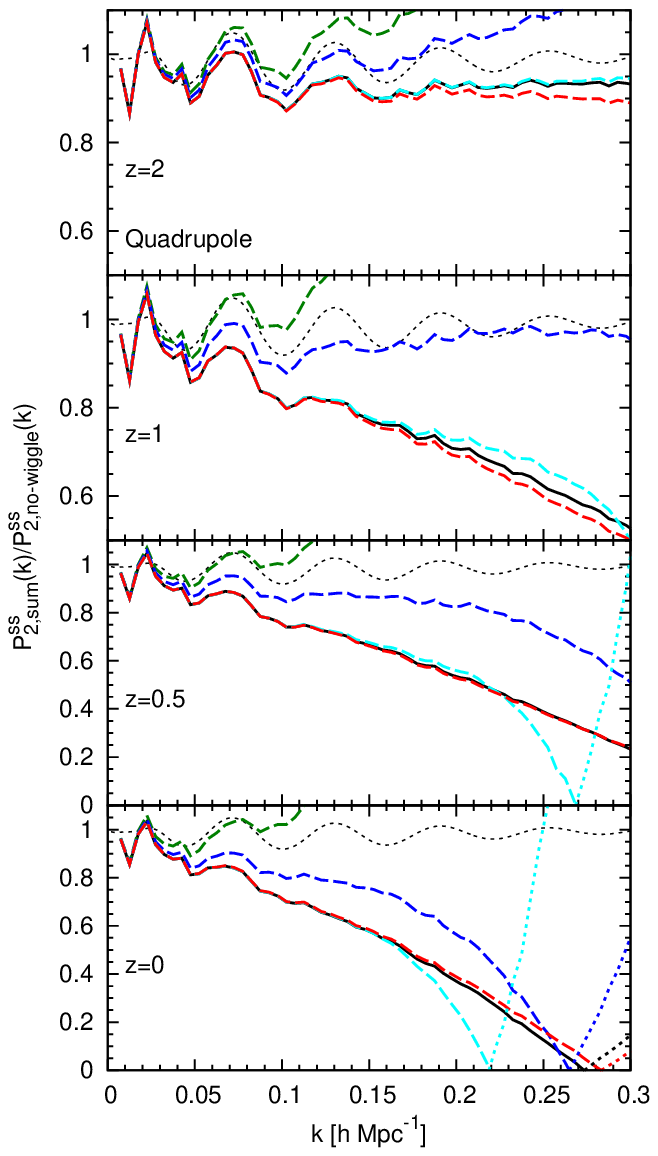}}
\caption{Redshift-space power spectrum divided by the no-wiggle
  approximation \cite{Eisenstein:1998}, monopole (left) and quadrupole
  (right).  The green, blue and cyan lines show our model prediction up to nonlinear 
  Kaiser, 2nd and 6th order corrections. The dotted black lines are linear
  theory prediction, while the solid black lines the reference power spectrum. 
  The red lines show our FoG model. The negative values of the quadrupole spectrum 
  on small scales are shown as the dotted lines. 
  }
\label{fig:pknw}
\end{figure}

In figure \ref{fig:pkl_mono} we show the monopole power spectrum at
$z=0$, 0.5, 1, and 2 summed up to nonlinear Kaiser, 2nd, 3rd, 4th, and
6th order approximations in $k\mu$ expansion. Here we denote the
summation at a given order as including all terms that have the same
$L+L'$: hence 2nd order includes all 3 Kaiser terms $P_{00}$, $P_{01}$
and $P_{11}$, as well as $P_{02}$, while the nonlinear Kaiser model
includes only the first three.  The lower panels show the error for a
given order, $P^{ss}_{0,{\rm sum}}/P^{ss}_{0,{\rm ref}}-1$.  The
linear theory power spectra with the input cosmological parameters of
our simulations are also plotted for comparison.  The power spectrum
of the nonlinear Kaiser model starts to deviate from the reference
spectrum at very large scales, $k\simeq 0.05 \hmpci$. However, adding
the term $P_{0 2}$, which has the same order contribution as $P_{11}$,
to the nonlinear Kaiser model, improves the accuracy.  Adding the
higher order terms continues to improve the accuracy down to smaller
and smaller scales, but the gains decrease as we approach the scale
$k=\sigma^{-1} \sim 0.2\hmpci$ ($z=0$), where the perturbative
expansion breaks down.  Our formula for the redshift-space monopole
spectrum $P^{ss}_{0,{\rm sum}}$, summed up to or more than 6th order,
is accurate within a few percent accuracy at $k\simeq 0.2 \hmpci$ at
$z=0$ and at $k\simeq 0.4 \hmpci$ at $z=2$.  It predicts not only the
overall shape of the redshift-space power spectrum up to these scales
but also baryon acoustic oscillations (BAO): to see this more clearly
we show the summed power spectra divided by the smoothed no-wiggle
spectrum \cite{Eisenstein:1998} in figure \ref{fig:pknw}.

%%%%%%%%%%%%%%%%%%%%%%%%%%%%%%%%%%%%%%%%%%%%%%
% Figure 2
%%%%%%%%%%%%%%%%%%%%%%%%%%%%%%%%%%%%%%%%%%%%%%
%%%%%%%%%%%%%%%%%%%%%%%%%%%%%%%%%%%%%%%%%%%%%%

%%%%%%%%%%%%%%%%%%%%%%%%%%%%%%%%%%%%%%%%%%%%%%
% Figure 3
%%%%%%%%%%%%%%%%%%%%%%%%%%%%%%%%%%%%%%%%%%%%%%

\begin{figure}%[!t]
\centering
\subfigure{\includegraphics[width=1.\textwidth]{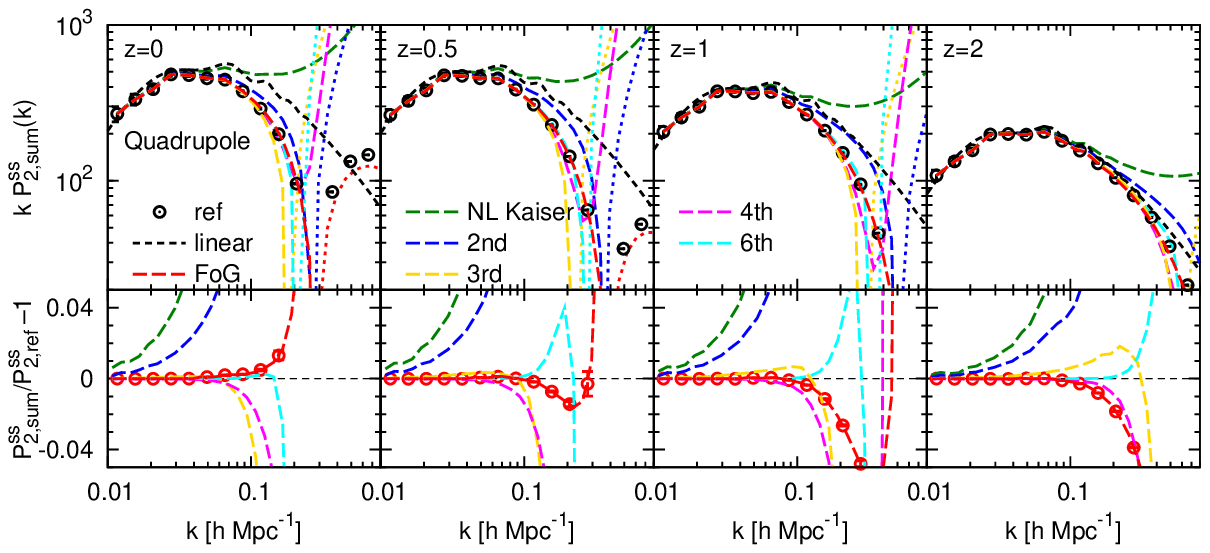}}
\caption{Same as figure \ref{fig:pkl_mono}, but for the
  quadrupole. The dashed lines at the top panels show positive values
  while the dotted lines show negative values.}
\label{fig:pkl_quad}
\end{figure}

\begin{figure}%[!t]
\centering
\subfigure{\includegraphics[width=1.\textwidth]{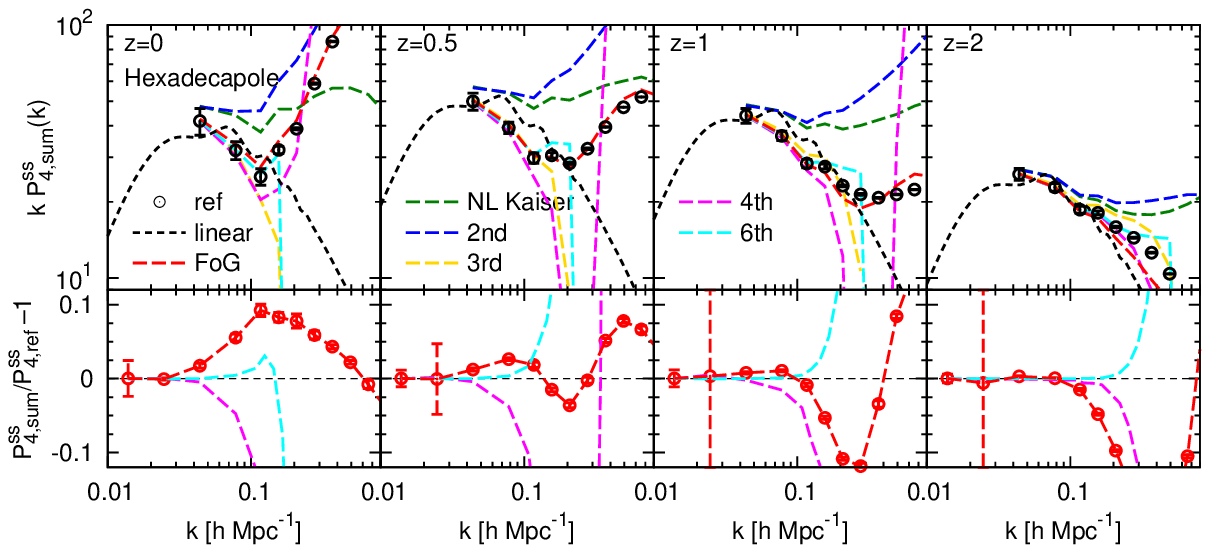}}
\caption{Same as figure \ref{fig:pkl_mono}, but for the hexadecapoles.
  We adopt the different bin sizes in logarithmic spacing at $k\leq
  0.1\hmpci$ and $k\geq 0.1\hmpci$.  Artificial cuts are put for the
  plots of the hexadecapoles at low $k$ because of large sampling
  variance.  }
\label{fig:pkl_hexa}
\end{figure}

Figure \ref{fig:pkl_quad} is the same as figure \ref{fig:pkl_mono},
but shows the results for the quadrupole spectra $P_2^{ss}$.  Because
the nonlinear quadrupole spectra crosses zero at high-$k$, there
exists a singularity point for the ratio of summed and reference
spectra at small scales.  The predictions for the quadrupole moment
reproduce the reference spectrum within a few percents up to the
scales of the singular point, $k\simeq 0.15 \hmpci$ at $z=0$ and
$k\simeq 0.3 \hmpci$ at $z=2$.  The quadrupole spectra divided by 
the corresponding no-wiggle spectrum are shown at the right side of 
Figure \ref{fig:pknw}. Figure \ref{fig:pkl_hexa} shows the
results for the hexadecapole spectrum.  We adopt broader $k$ binning
for the hexadecapole moment at $k<0.1\hmpci$ and put artificial cuts
for the plots of $kP^{ss}_{4}$ because of large sampling variance.  We
do not show the results obtained from the nonlinear Kaiser, 2nd and
3rd order approximations in lower panels because they strongly deviate
from the reference power spectrum (as shown in the upper panels of
figure \ref{fig:pkl_hexa}).  Although the measurement of the
hexadecapole moment of the redshift-space power spectrum is very
noisy, the higher order expansion predictions give a good agreement if
we consider summation to 6th order.

%%%%%%%%%%%%%%%%%%%%%%%%%%%%%%%%%%%%%%%%%%%%%%
% Figure 4
%%%%%%%%%%%%%%%%%%%%%%%%%%%%%%%%%%%%%%%%%%%%%%

%%%%%%%%%%%%%%%%%%%%%%%%%%%%%%%%%%%%%%%%%%%%%%
% Figure 5
%%%%%%%%%%%%%%%%%%%%%%%%%%%%%%%%%%%%%%%%%%%%%%

%%%%%%%%%%%%%%%%%%%%%%%%%%%%%%%%%%%%%%%%%%%%%%
% Section 4 FoG resummation
%%%%%%%%%%%%%%%%%%%%%%%%%%%%%%%%%%%%%%%%%%%%%%
\section{Fingers-of-God resummation}\label{sec:fog_resum}
It is clear from the results in previous section that while for $k\mu
\sigma/H <1$ we have a convergence and only a finite number of terms
need to be considered, there is no convergence for $k\mu \sigma/H >1$:
individual terms become larger and larger as we go to higher orders,
yet the total sum in $P^{ss}(k,\mu)$ remains well behaved.  This
suggests we need to explore ways to resum the terms.

While it is difficult to make progress in general terms, there are
specific situations that can be controlled.  We are interested in a
situation where pieces of terms disconnect within the correlation
function. For example, in $P_{02}$ term we correlate $\delta$ with
$(1+\delta)u_{\parallel}^2$, and if $u_{\parallel}^2$ is dominated by
the small scales, as will be the case in systems with large velocity
dispersion caused by nonlinear gravitational collapse, then on large
scales this term becomes $P_{02} \sim P_{00}\sigma^2$, where
$\sigma^2=\langle u_{\parallel}^2\rangle$ and the total contribution
to $P^{ss}$ is $-P_{00}(k\mu\sigma/H)^2$. We see that this term scales
as the linear order term, but has opposite sign to it, i.e. This term
suppresses power and this suppression scales as $k^2$.  The long range
correlation is entirely in the density field.  While this analysis
suggests the mean square velocity field enters as the physical
parameter, as discussed in \cite{Seljak:2011}, any bulk velocity that
displaces particles as a solid body will not contribute to FoG. This
cancellation shows up in part of $P_{11}$ cancelling $P_{02}$, such
that only the dispersion part enters into the $\sigma^2$ term, while
the bulk part cancels.

One can identify similar terms at higher order, for example
$P_{04}\sim P_{00}\langle u_{\parallel}^4\rangle$ and one can write
$\langle u_{\parallel}^4\rangle=3\sigma^4+\langle
u_{\parallel}^4\rangle_c$, where $\langle u_{\parallel}^4\rangle_c$ is
the connected part of the curtosis. There will also be a term
$P_{22}\sim P_{00} \sigma^4$.  The total contribution to $P^{ss}$ from
these terms is thus
\begin{eqnarray}
P^{ss}(k,\mu)&=& P_{00}[1-(k\mu \sigma/H)^2+(k\mu \sigma/H)^4/2+2(k\mu/H)^4 \langle u_{\parallel}^4\rangle_c/4!\cdots] \nonumber \\
&=& P_{00}e^{-(k\sigma\mu/H)^2+2(k\mu/H)^4\langle u_{\parallel}^4\rangle_c/4!\cdots}. \label{eq:p_ss_resum}
\end{eqnarray}
The same calculation can be done at the field level. 
We are interested in the situation where the pieces of a term like 
$\delta(\vx)u_\parallel^2(\vx)$ are disconnected within a correlation calculation,
e.g., $\left\langle X(\vy)\delta(\vx)u_\parallel^2(\vx) \right\rangle = 
\left\langle X(\vy)\delta(\vx)\right\rangle \left\langle u_\parallel^2\right\rangle +$
other terms. We can re-sum the pieces into a FoG factor
\be
G^{1/2}(k\mu) =\exp{\left[ \sum_{L=1} \frac{1}{L!}\left( \frac{ik\mu}{H} \right)^L \langle u_{\parallel}^L \rangle_c \right]},
\ee
where $\langle u_\parallel^L \rangle_c$ is the connected part of $\langle u_\parallel^L \rangle$ and note that 
the odd $L$ terms are zero by symmetry. The lowest order term here is just the usual Gaussian kernel 
$\exp\left( -k^2\mu^2 \sigma^2/2H^2 \right)$. 
After this re-summation we can write the redshift-space density field as
\be
\delta_s(\vk)=G^{1/2}(k\mu) \left[ \delta(\vk) + \sum_{L=1}\frac{1}{L!} \left(\frac{ik\mu}{H}\right)^L 
\left[ \left(1+\delta(\vx)\right) \left[ u_\parallel^L\right]_c(\vx)  \right]_\vk \right],
\ee
where $\left[ u_\parallel^L\right]_c$ is understood to be $u_\parallel^L$ minus all possible internal 
averages of any number of $u$'s.
This motivates us to write
\begin{equation}
P^{ss}(\vk)=G([k\mu\sigma/H]^2)P_{\rm Kaiser}(\vk),
\end{equation}
where $P_{\rm Kaiser}$ account for the lowest 3 terms given by
equation (\ref{eq:nl_kaiser}) and where $G(x)$ is exponential in the
simplest case where higher order reduced moments can be ignored, while
more generally it is a function with alternating signs of coefficients
in Taylor expansion. We have written $G(x)$ in terms of $x=[k\mu
  \sigma/H]^2$ only: if curtosis is present then we can either write
it by adding additional $(k\mu)^4$ terms to the exponential,
$\exp[-(k\sigma\mu/H)^2+2(k\mu/H)^4\langle
  u_{\parallel}^4\rangle_c/4!\cdots]$, or, equivalently, we allow for a
more general functional form of $G(x)$ than an exponential.

Note that in the simplest form this ``derivation" gives exactly the
exponential FoG form proposed in the literature \cite{Peacock:1994,
  Park:1994, Ballinger:1996, Scoccimarro:2004, Percival:2009}.  Other
forms for $G(x)$ have been proposed in the literature, e.g.
\begin{eqnarray}
G(x=(k\mu\sigma/H)^2)=\left\{ 
\begin{array}{ll}
\left( 1+ x \right)^{-1} & {\rm Lorentzian}, \\
\exp{\left( -x \right)} & {\rm Gaussian},
\end{array}\right. \label{eq:fog_phenom}
\end{eqnarray}
where $\sigma$ was treated as a free parameter.  See \cite{Tang:2011,
  Kwan:2011} for the studies to adopt more than one free parameter for
the FoG term.  These empirical forms all behave qualitively in the
same manner, i.e. to first order of Taylor expansion in $(k\mu) ^2$
they give $G=1-(k\mu\sigma/H)^2+ {\cal O}([k\mu \sigma/H]^4)$ and
higher order terms alternate in the signs.

The effective velocity dispersions can be evaluated from our
expression of the redshift-space power spectrum (equation
(\ref{eq:p_ss})) and uniquely resummed into the FoG kernels because the
higher-order terms of the expansion contain all information.  We know
that a Taylor series of the FoG kernels like in equation
(\ref{eq:fog_phenom}) produce positive and negative terms alternatively,
just like the terms in equation (\ref{eq:p_ss}).  
However, there are a lot of $P_{LL'}$ terms in our expansion. 
At each even order both positive and negative terms appear, 
e.g., at 4th order the $P_{04}$ and $P_{22}$ terms have positive contributions while the $P_{13}$ 
term has a negative contribution. 
On the other hand, at odd order contributions are all positive and negative in tern, e.g., 
at 3rd order the terms are all negative and at 5th order all positive. 
Thus we decided to include 
three independent FoG kernels for the two even linear order order terms ($P_{00}$ and $P_{11}$)
and one for the lowest odd order $P_{0 1}$.  These three terms thus need to be
individually multiplied by something which corresponds to the
generalized FoG kernels. To do this, we can apply the same reasoning as the discussion around 
equation (\ref{eq:p_ss_resum}) to other terms that contain long range correlations between density 
and velocity, i.e., portions of terms such as $P_{03}$ and $P_{13}$ can be respectively written 
as $P_{03}\sim -6P_{01}(k\mu\sigma/H)^2$ and $P_{13}\sim -3P_{01}(k\mu\sigma/H)^2$, which,
together with the higher order terms, can be resummed into a similar function of even orders in powers of $k\mu/H$. 

Finally, we can write
\begin{equation}
  P^{ss}(\vk)=G_{00}\left( [k\mu \sigma_{00}/H]^2\right)P_{00} 
  + 2G_{01}\left( [k\mu \sigma_{01}/H]^2 \right)\frac{ik\mu}{ H}P_{01}
  + G_{11}\left( [k\mu \sigma_{11}/H]^2 \right)\frac{(k\mu)^2}{ H^2}P_{11}.  \label{eq:p_ss_fog}
\end{equation}
We have defined three different velocity dispersions $\sigma_{00}$,
$\sigma_{01}$, and $\sigma_{11}$.  The expressions of the velocity
dispersions in the FoG kernels can be uniquely derived from equation
(\ref{eq:p_ss}) as
\begin{eqnarray}
  \sigma_{00}^2 (\vk) &=&\frac{P_{02}}{P_{00}}, \label{eq:sig_ss_00} \\
  \sigma_{0 1}^2(\vk)&=&\frac{1}{6}\frac{P_{03}+3P_{21}}{P_{01}}, \\
  \sigma_{11}^2(\vk)&=&\frac{1}{3}\frac{P_{13}}{P_{11}}. \label{eq:sig_ss_11}
\end{eqnarray}

To go beyond that and determine the form of FoG kernel we expand the
FoG terms as a Taylor series in terms of $(k\mu)^2$.  Because all the
phenomenological FoG models have the first derivative equal to $-1$,
we define $dG_{LL'}(x)/dx |_{x=0}=-1$.  Now let us consider the ansatz
for the FoG model,
\begin{equation}
  G_{LL'} \left(x_{LL'}; \alpha_{LL'} \right)=\left(1+ \frac{x_{LL'}}{\alpha_{LL'}} 
  \right)^{-\alpha_{LL'}}~,  \label{eq:fog}
\end{equation}
where $x_{LL'}=(k\mu\sigma_{LL'}/H)^2$.  Each FoG kernels contains two
parameters, $\sigma_{LL'}$ and $\alpha_{LL'}$. These can reproduce the
functional forms in previous studies: Lorentzian for $\alpha_{LL'}=1$
and Gaussian for $\alpha_{LL'}=\infty$.  The FoG parameter
$\alpha_{LL'}$ is related to $n$-th derivative of $G_{LL'}$ as
$(n\geq2)$
\begin{equation}
  \frac{d^n}{dx^n}G_{LL'}(x)|_{x=0}=
  (-1)^n\prod^{n-1}_{m=1} \left(1+\frac{m}{\alpha_{LL'}}\right)~. \label{eq:kernel_derivative}
\end{equation}
We use the $\alpha_{LL'}$'s determined from the lowest contribution at
second order, as
\begin{equation}
	\alpha_{LL'}(\vk)=\left[ G_{LL'}^{''}(\vk)-1\right]^{-1} . \label{eq:alpha_def}
\end{equation}
Note that the expressions for our FoG models of equation
(\ref{eq:p_ss_fog}) preserve full generality up to order $(k\mu/H)^5$. 
The quality of the ansatz can be investigated by looking at higher
order terms: the comparison of the derivatives of the FoG kernels up
to 4th order with the expansion terms is given in appendix
\ref{sec:derivative}.

The FoG model in equation (\ref{eq:fog}) is not the only option to
choose.  As discussed above, the resummation can be expressed more
elegantly in terms of connected moments, where the cumulant theorem
naturally leads to an expression of the form 
\be G_{LL'}=\exp[-(k\mu
  \sigma/H)^2+(k\mu\tau/H)^4+\cdots], \label{eq:kurtosis} 
\ee 
where the lowest two cumulants are variance $\sigma^2=\langle
u_\parallel^2 \rangle$ and kurtosis $\tau^4=\langle u_\parallel ^4
\rangle _c$, which is the connected part of $\langle u_\parallel ^4
\rangle$.  We can generalize this expression and introduce 3 variance
and 3 curtosis terms separately for 00, 01, and 11 terms, and we can
relate the kurtosis terms $\tau_{LL'}$'s to the parameters
$\alpha_{LL'}$'s defined above as
$2(\tau_{LL'}/\sigma_{LL'})^4=1/\alpha_{LL'}$.

Let us summarize this discussion: the decoupling of small scale
velocity dispersion like terms from the long range correlations
motivates a resummation of the terms into the so called FoG kernels,
which multiply the long range correlation terms contained in
density-density ($P_{00}$), density-momentum ($P_{01}$) and
momentum-momentum ($P_{11}$) correlators. Only a portion of the terms
can be motivated in such a way, while other terms are simply nonlinear
couplings that do not reduce to linear order correlation on large
scales.  It is therefore difficult to provide a formal justification
for this resummation, but it is worth analyzing to what extent this
approach is useful. Here we use the distribution function expansion in
equation (\ref{eq:p_ss}) to formally define FoG kernel parameters: up to
$(k\mu)^5$ 
this is equivalent to the exact expansion, but requires 3
different FoG kernels. An additional point that needs to be emphasized
is that these quantities as we defined are a function of angle and
scale, i.e. they are $\mu$ and $k$ dependent.  Below we use numerical
simulations to compare the original expansion in equation
(\ref{eq:p_ss}) to the FoG resummation version to see to what extent FoG
approach is useful for the general treatment of RSD.

\subsection{Testing the Fingers-of-God model}\label{sec:fog}

\begin{figure}[!t]
\subfigure{\includegraphics[width=1.\textwidth]{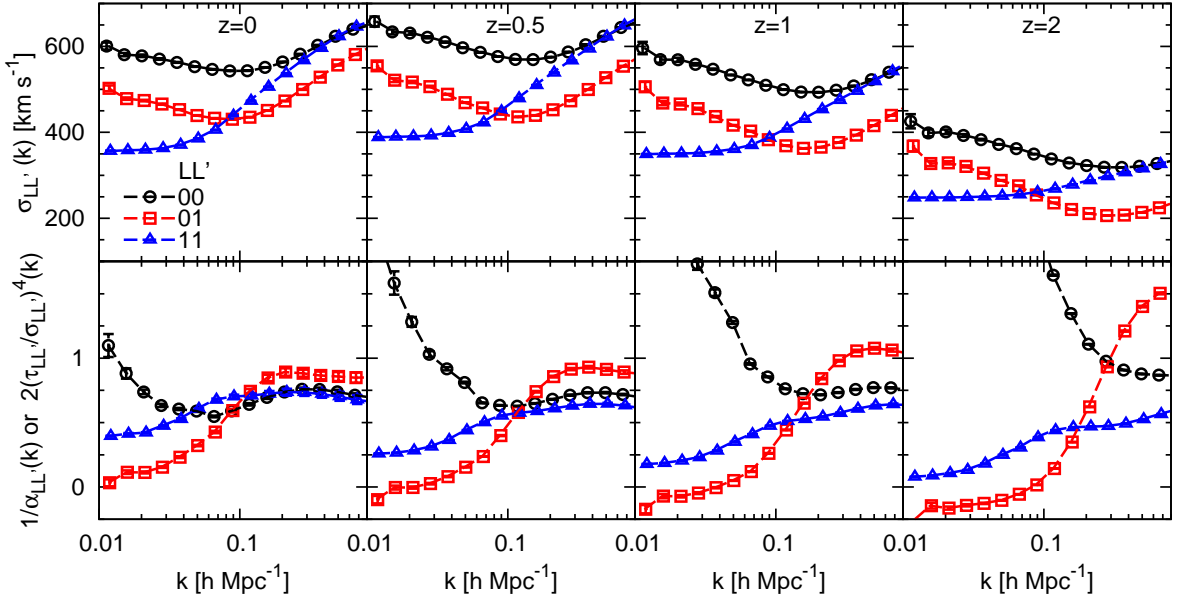}}
\caption{Spherically averaged FoG parameters, velocity dispersions
  (upper panels) and the power law index of FoG kernel (lower panels).   
  The circles/triangles have been respectively offset in the negative/positive direction
  for clarity.
  }
\label{fig:sigvk}
\end{figure}

In this subsection we compare the FoG resummation of the higher order
terms discussed above to numerical simulations.  The upper panels of
figure \ref{fig:sigvk} show velocity dispersions determined from our
simulations using equations (\ref{eq:sig_ss_00}) -- (\ref{eq:sig_ss_11}).
For clarity we plot the spherically-averaged velocity dispersions,
i.e., those obtained from the monopole spectra.  While the 3 velocity
dispersion terms are similar to each other, they differ in the
amplitude and scale dependence, suggesting that it is important to
independently consider the FoG kernel for each spectrum in the
nonlinear Kaiser formula.  Note that these values are higher than
those determined by previous studies (e.g., figure 7 of
\cite{Taruya:2010}).  This is because our quantities are based on the
mass-weighted velocities in contrast to the volume-weighted velocities
discussed in \cite{Taruya:2010}.  We discuss the latter in section
\ref{sec:v-expansion}.

%%%%%%%%%%%%%%%%%%%%%%%%%%%%%%%%%%%%%%%%%%%%%%
% Figure 6
%%%%%%%%%%%%%%%%%%%%%%%%%%%%%%%%%%%%%%%%%%%%%%
%%%%%%%%%%%%%%%%%%%%%%%%%%%%%%%%%%%%%%%%%%%%%%

\begin{figure}[!t]
\centering
\includegraphics[width=.8\textwidth]{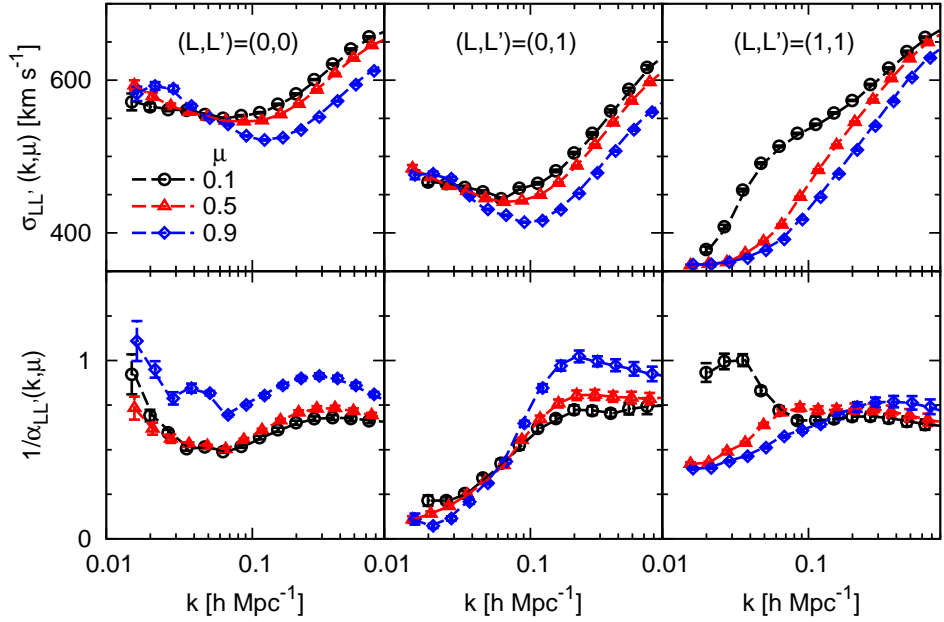}
\caption{Angular dependence of FoG parameters at $z=0$. 
The width of $\mu$ bin is 0.2 centered around the values shown in the top left panel.}
\label{fig:sigv}
\end{figure}

The lower panels in figure \ref{fig:sigvk} show the inverses of the
second parameters for our FoG model, $1/\alpha_{LL'}(k)$ defined by
equation (\ref{eq:alpha_def}).  As we have discussed above, the
parameters are equivalent to $2\left( \tau_{LL'}/\sigma_{LL'} \right)^4$
when we adopt the kurtosis as the second parameters.  Note that for a
Lorentzian model and a Gaussian model we have $1/\alpha_{LL'}=1$ and
$1/\alpha_{LL'}=0$, respectively.  The impression from the figure is
that $\alpha$ parameters have strong $k$ dependences and behave
differently from each other at large scales.  However, the FoG terms
have negligibly small effects on the shape of the power spectrum at
large scales.  On the other hand, at lower redshifts the $\alpha$
values converge to nearly a constant at $k>0.1\hmpci$ where the FoG
effect starts to play an important role, with typical values between
$1<\alpha_{LL'}<2$, i.e. Lorentzian FoG kernel is a better
approximation than Gaussian.  The convergence of $\alpha_{LL'}$ to a
single value is worse as we go to higher redshifts and the difference
between the values of $\alpha_{LL'}$'s remains large.  In order to see
the angular dependence of our FoG parameters, we show $\sigma_{LL'}$
and $1/\alpha_{LL'}$ at $z=0$ as functions of $k$ and $\mu$ at the
upper and lower panels in figure \ref{fig:sigv}, respectively.

%%%%%%%%%%%%%%%%%%%%%%%%%%%%%%%%%%%%%%%%%%%%%%
% Figure 7
%%%%%%%%%%%%%%%%%%%%%%%%%%%%%%%%%%%%%%%%%%%%%%
%%%%%%%%%%%%%%%%%%%%%%%%%%%%%%%%%%%%%%%%%%%%%%

Now we focus on the power spectrum $P^{ss}$ using the FoG model
discussed above.  In figures \ref{fig:pkl_mono}, \ref{fig:pkl_quad}
and \ref{fig:pkl_hexa} the resulting power spectra with the FoG
kernels are shown as the red points.  As we have seen in section
\ref{sec:power_measure}, there were certain scales at which the power
spectrum from our perturbative expansion approach breaks down and
diverse, $k\simeq 0.2 \hmpci$ at $z=0$ and $k\simeq 0.4 \hmpci$ at
$z=2$ for the monopole spectrum even though we sum up to the 6th order
terms.  Our FoG model dramatically improves the results.  The accuracy
of a few percent is achieved up to $k\simeq 0.4 \hmpci$ at $z=0$ and
even up to to $k <1.0 \hmpci$ at $z=0.5$ and 1. On the other hand, the
improvement at $z=2$ is not so much better than the lower redshfits
because of a strong suppression by the FoG.  Our FoG model works for
the quadrupole spectrum as well as for the monopole spectrum: it
predicts the quadrupole spectrum down to very small scales (see the
upper panels of figure \ref{fig:pkl_quad} and the right panels of figure \ref{fig:pknw}).  
Even the accuracy of the hexadecapole spectrum of our FoG model is quite good, 
at the level of 10\%, and it reproduces the shape over all the scales probed (see
upper panel of figure \ref{fig:pkl_hexa}).

\subsection{Mass-weighted vs volume-weighted velocities} \label{sec:v-expansion}

It is worth comparing our approach based on the power spectra of
mass-weighted velocity moments with those of volume-weighted velocity
moments, which are commonly used to model nonlinear power spectra
(e.g., \cite{Scoccimarro:2004, Percival:2009, Taruya:2010,
  Jennings:2011, Tang:2011}).  The difference comes from the fact that
the $L$-th moment of the mass-weighted moments
$T_\parallel^L=(1+\delta)u_{\parallel}^L$ in equation (\ref{eq:deltak})
contains contributions not only from $L$-th order in perturbation
theory, but also $(L+1)$-th order because of the term
$\delta(\vx)u^L_\parallel(\vx)$.  By regrouping equation
(\ref{eq:deltak}) with the same order term, we obtain
\begin{equation}
  \delta_s(k,\mu) = \delta(k,\mu)+
  \frac{ik_\parallel}{\mathcal{H}}u_\parallel(k,\mu) + \sum_{L=2}^{\infty} \left[
    \frac{1}{(L-1)!} \left(\frac{i
      k_\parallel}{\mathcal{H}}\right)^{L-1}
    \left[\delta(\vx)u_\parallel^{L-1}(\vx)\right]_\vk + \frac{1}{L!}
    \left(\frac{i k_\parallel}{\mathcal{H}}\right)^{L}
    u_\parallel^{L}(\vk) \right]~, \label{eq:deltak_ex_v}
\end{equation}
where the first and the second terms of the right-hand side are
respectively the zeroth and first order terms, while the bracketed
terms correspond to the $L$-th order terms $(L\geq 2)$.  By squaring
equation (\ref{eq:deltak_ex_v}), we obtain the same equation as equation
(\ref{eq:p_ss}), but the terms regrouped into the same order in SPT,
\begin{eqnarray}
  P^{ss}(\vk)=P_{00} + 2
  \frac{ik\mu}{ H} P_{0 u^1_{\parallel}} + 
  \frac{(k\mu)^2}{ H^2} P_{u^1_{\parallel}u^1_{\parallel}} 
  + \left[ 2 \frac{ik\mu }{ H}
    P_{0 w^1_{\parallel}} 
  - \frac{(k\mu)^2 }{ H^2}  P_{0 u^2_{\parallel}} \right] + \cdots
 \label{eq:pkmu_v}
\end{eqnarray}
where $w^L_{\parallel}=T_\parallel^L-u_\parallel^L=[\delta
  u_\parallel^L]_{\vk}$, thus $P_{0
  w^L_{\parallel}}=P_{0L}-P_{0u_{\parallel}^L}$ and similarly for
higher orders.

As many previous studies have already discussed
(e.g. \cite{Tang:2011}), measuring the power spectrum of
volume-weighted velocity moments is not as straight-forward as
measuring mass-weighted velocity moments used in our formalism.  In
order to measure the moments of volume-weighted velocities, we divide
the interpolated moments of mass-weighted velocities by the
interpolated density before the field is Fourier-transformed
$T_\parallel^L/(1+\delta)=u^L_\parallel$.  This can be noisy: some
points on the grid may not have any particles, so a sufficiently
coarse grid is needed. More importantly, the results depend on the
grid size, i.e. on smoothing (see \cite{Scoccimarro:2004,
  Pueblas:2009} for a detailed discussion of how to measure the
volume-weighted velocities from $N$-body simulations).  Following the
same process as described in section \ref{sec:nbody}, we compute the
redshift-space power spectrum based on the power spectra of
volume-weighted velocity moments up to a given level of accuracy.

\begin{figure}%[!t]
\centering
\subfigure{\includegraphics[width=1.\textwidth]{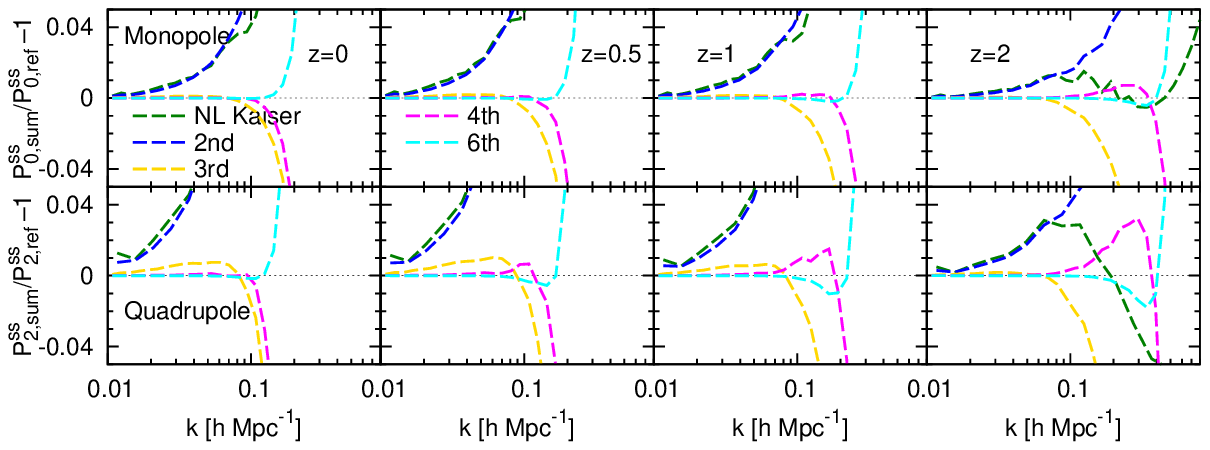}}
\subfigure{\includegraphics[width=1.\textwidth]{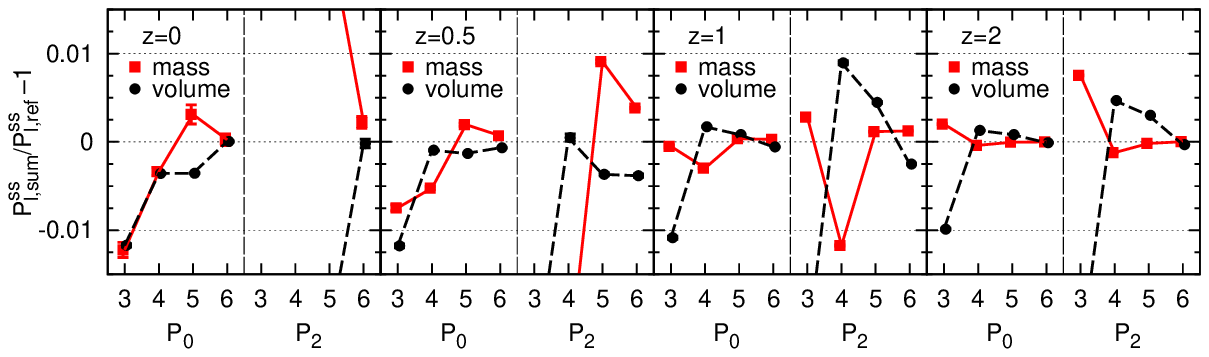}}
\caption{({\it Top set}) Error between the summed power spectrum in terms of volume weighted velocity moment expansion and the reference spectrum. Lines are the same as the lower panels of figures
  \ref{fig:pkl_mono} and \ref{fig:pkl_quad}.  
  ({\it Bottom set}) Error between the summed power spectrum and the reference spectrum at $k=0.117\hmpci$. 
The left and right sides of each panel show the monopole and quadrupole spectra, respectively. 
The horizontal axis is a given order in terms of mass-weighted moment expansion (red points) and 
volume-weighted moment expansion (black points).
}
\label{fig:pkl_ratio_v}
\end{figure}

The top set of figure \ref{fig:pkl_ratio_v} shows the ratio of the summed power
spectra of equation (\ref{eq:pkmu_v}) to the corresponding reference
spectra in redshift space. Because equation (\ref{eq:pkmu_v}) is
essentially the same as equation (\ref{eq:p_ss}), here we want to see
the convergence of these expressions.  At the order of first 3
(Kaiser) terms, the expansion with volume-weighted velocities is
somewhat closer to the reference spectrum than that with mass-weighted
velocities, but both approximations are bad.  
The bottom set of figure \ref{fig:pkl_ratio_v} shows the same quantities as in the top set but 
those at $k=0.117\hmpci$ at from 3rd to 6th orders. 
At higher orders the
convergence is slightly faster with mass-weighted quantities for monopoles when they 
are compared at the same order: one can
predict well the redshift-space power spectrum at $k=0.1\hmpci$ by
including the 4th and 3rd order corrections at $z=0$ and $z>0$,
respectively, for our power spectrum of mass-weighted velocity
moments, while one needs to include the 4th order corrections 
for the power spectrum of volume-weighted velocity moments. 
For the quadrupole spectra, the situation is more complicated but the value of 
the spectra from the volume-weighted velocity moments expansion is deviated 
from the reference value at 3rd order at very large scales as seen at the 
bottom panels of the top set in figure \ref{fig:pkl_ratio_v}. Both approaches
break down once one enters the non-perturbative regime ($k>0.2\hmpci$
at $z=0$).  We conclude that in terms of rate of convergence, there is
no advantage in defining volume weighted quantities.  If one works
with galaxies and halos number density weighting becomes essential,
since it is difficult to define volume weighted velocity moments in a
sparsely sampled system \cite{Tang:2011}.  In that situation our
approach is the most meaningful way to define physical quantities that
enter in RSD description.

%%%%%%%%%%%%%%%%%%%%%%%%%%%%%%%%%%%%%%%%%%%%%%
% Figure 8
%%%%%%%%%%%%%%%%%%%%%%%%%%%%%%%%%%%%%%%%%%%%%%
%%%%%%%%%%%%%%%%%%%%%%%%%%%%%%%%%%%%%%%%%%%%%%

\section{Expansion in powers of $\mu^2$}\label{sec:angular}

As we have seen in previous two sections the series expansion of
equation (\ref{eq:p_ss}) is convergent on large scales, but not on small
scales. For sufficiently high $k$ any finite order summation fails
drastically.  FoG resummation approach fares better, but even that
fails for high $k$. One can sidestep these issues by considering an
alternative expansion in terms of powers of $\mu^2$: as discussed in
section \ref{sec:th_angular} and in \cite{Seljak:2011} for any finite
power of $\mu^2$ there is a finite number of $P_{LL'}$ terms
contributing to it.  For $\mu^0$ only $P_{00}$ contributes, for
$\mu^2$ $P_{01}$, $P_{11}$ and $P_{02}$, for $\mu^4$ $P_{11}$,
$P_{02}$, $P_{03}$, $P_{12}$, $P_{04}$, $P_{13}$ and $P_{22}$ etc.

Only these 3 lowest terms, $\mu^0$, $\mu^2$, and $\mu^4$, contain
cosmological information at the linear order, so in principle these
are the only relevant terms.  However, if we expand the full
$P^{ss}(k,\mu)$ into powers of $\mu^2$ and try to determine the
coefficients from the data, the resulting coefficients will be
correlated: only Legendre expansion assures uncorrelated values. As a
result there will be mixing of higher powers of $\mu^2$ into lower
powers if they are not accounted for in the fits, or there will be
strong degeneracies and the fits will be unstable if all the
coefficients are accounted for but we allow them to take any value.
Typically one solves this by regularizing the expansion, i.e. by
constraining them to a certain range of values.  In this paper we will
not focus on methods how to determine the coefficients of such
expansion from the data, but we will show $\mu^6$ and $\mu^8$
expansion terms to develop some understanding of their scale
dependence and amplitude.

\subsection{$\mu^2$ terms}

\begin{figure}%[!t]
\centering
\subfigure{\includegraphics[width=1.\textwidth]{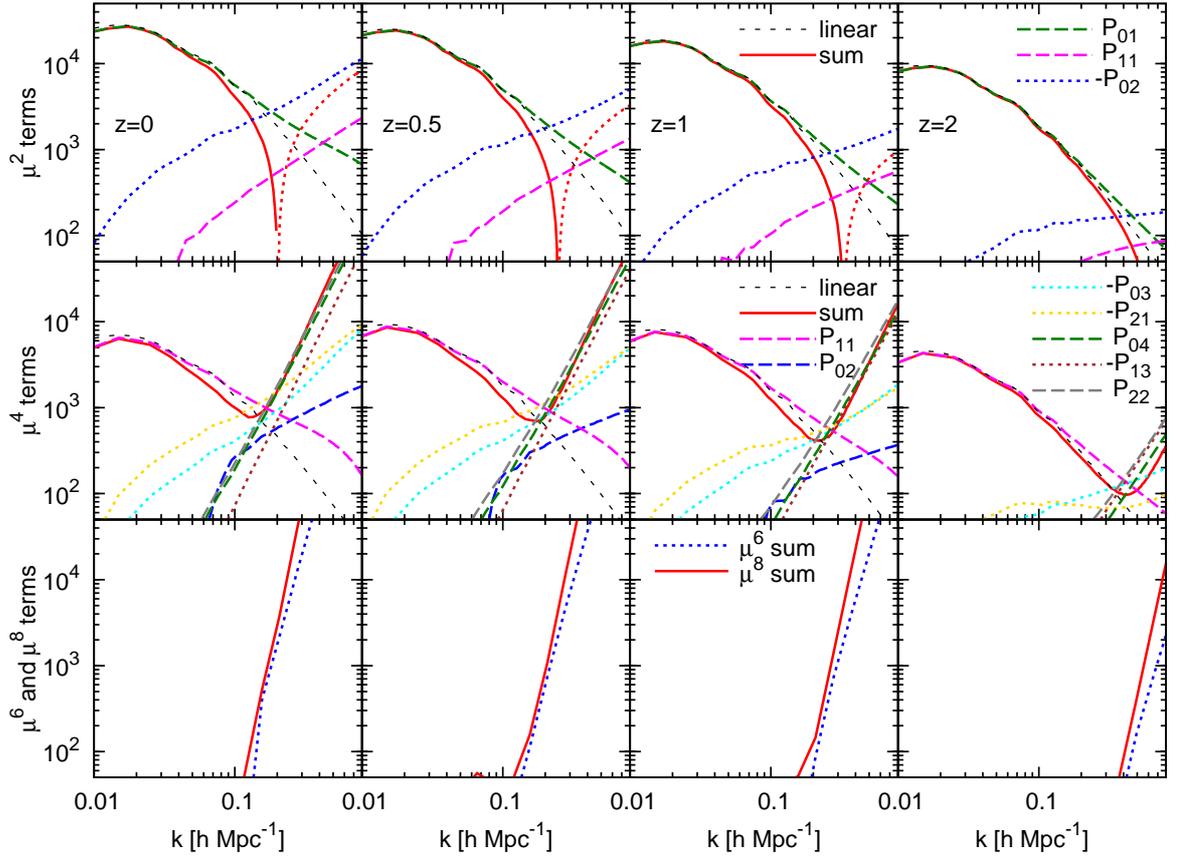}}
\caption{Contributions to $\mu^2$ (upper panels), $\mu^4$ (middle
  panels), and $\mu^6$ and $\mu^8$ terms (lower panels) in
  redshift-space power spectrum $P^{ss}(k,\mu)$.  The positive
  contributions are shown as the solid and long-dashed lines while the
  negative contributions as the dotted lines (linear theory
  predictions are shown as the short-dashed black line).  }
\label{fig:mu}
\end{figure}

In the top panels of figure \ref{fig:mu} we show these individual term
contributions to the lowest order powers of $\mu$ (we do not show
$\mu^0$ term, which is just the usual real space power spectrum
$P_{00}$).  For $\mu^2$ we see that the $P_{01}$ dominates for low
$k$, as expected since that is the only term that does not vanish in
linear theory. This term follows linear theory prediction for low $k$,
while for $k>0.1\hmpci$ it exceeds it, just like it happens for the
dark matter power spectrum $P_{00}$ itself.  This is not surprising:
as shown in \cite{Seljak:2011} we can write $ P_{01}^{ss}(\mu^2)={d
  P_{00}(k) \over d\ln a}$, so this term is given by the time
derivative of the dark matter power spectrum and has a similar scale
dependence relative to the linear power spectrum.

%%%%%%%%%%%%%%%%%%%%%%%%%%%%%%%%%%%%%%%%%%%%%%
%Figure 9
%%%%%%%%%%%%%%%%%%%%%%%%%%%%%%%%%%%%%%%%%%%%%%
%%%%%%%%%%%%%%%%%%%%%%%%%%%%%%%%%%%%%%%%%%%%%%

The next term in terms of relevance is $P_{02}$.  This term has
contributions from the correlation between the energy density and the
density $P_{00}^{020}$, as well as from the scalar part of the
anisotropic stress correlated with the density $P_{02}^{020}$, both
helicity 0 scalars.  As discussed in section \ref{sec:fog_resum}, we
expect the first term to be dominant and scale as
$-k^2P_{00}\sigma^2/H^2$, hence to dominate over $P_{01}$ for
$k\sigma/H>1$. This gives $\sigma=500{\rm km/s}$ at $z=0$, decreasing
to 200km/s for $z=2$. As expected we see this term is always negative.

The third term contributing to $\mu^2$ is the vector part (helicity 1)
of auto-correlation of momentum density with itself $P_{11}^{111}$.
This term is always positive and partially cancels $P_{02}$. As
discussed in \cite{Seljak:2011}, it cancels all the bulk motion
contributions to $\sigma^2$. While this term scales in a similar way
as $P_{02}$, it is 3-10 times lower in amplitude, so it cancels only a
small part of $P_{02}$.

We see that the total sum never exceeds the linear power spectrum and
becomes negative for $k \sim 0.17\hmpci$ at $z=0$ and $k \sim
0.5\hmpci$ at $z=2$. For scales smaller than that the $\mu^2$ term is
negative as a consequence of a strong FoG effect.

\subsection{$\mu^4$ terms}

This term receives contributions from 7 different terms, $P_{11}$,
$P_{02}$, $P_{03}$, $P_{12}$, $P_{04}$, $P_{13}$ and $P_{22}$. They
are shown at the middle panels in figure \ref{fig:mu}.  On large
scales the dominant term is $P_{11}$ which contains a linear order
contribution.  This term agrees with linear theory prediction for
$k<0.1\hmpci$ and is above that for $k>0.1\hmpci$, just like in the
case of $P_{00}$ and $P_{01}$.

The next order term in significance should be $P_{02}$.  We see this
term is relatively small and does not dominate anywhere.  $P_{02}$
contribution to $\mu^4$ arises entirely from from the scalar part of
the anisotropic stress correlated with the density $P_{02}^{020}$,
which contains $\mu^4$ term contribution to $P^{ss}$, while
$P_{00}^{020}$ does not. Physically one expects the small scale
velocity dispersion to be relatively isotropic, hence anisotropic
stress density should be small compared to the energy density.

Next two terms in terms of $L+L'$ are $P_{03}$ and $P_{12}$. These two
terms provide the dominant correction to $P_{11}$ on intermediate and
large scales $k<0.1\hmpci$. On very small scales terms $P_{04}$,
$P_{22}$ and $P_{13}$ dominate. As discussed in FoG section we expect
the first two terms to scale as $3k^4\sigma^4P_{00}$ and be equal in
amplitude, while $P_{13}$ should be negative and cancel out the bulk
flow part of the other two terms. We see that this expectation is
borne out in simulations: in total these terms add power on small
scales. The transition happens at a similar scale as for $\mu^2$ term,
$k \sim 0.17\hmpci$ at $z=0$ and $k \sim 0.5\hmpci$ at $z=2$.

\subsection{Higher order terms}

At order higher than $\mu^4$ we do not have any linear order
contributions, so we expect these terms to be small on large
scales. There are many terms that contribute. At the bottom panels in
figure \ref{fig:mu} we show the total contributions to $\mu^6$ and
$\mu^8$ terms. We can see that these terms are indeed negligibly small
at $k< 0.1\hmpci$ at $z=0$ and $k< 0.3 \hmpci$ at $z=2$.  At smaller
scales these contributions increase with the scale dependences of
$k^6$ and $k^8$, respectively.

\section{Conclusions} \label{sec:conclusion}
In this paper we used numerical simulations to investigate the
distribution function expansion approach to the redshift space
distortion power spectrum \cite{Seljak:2011}.  The power spectrum in
redshift space can be written as a sum over correlators between
mass-weighted velocity moments.  We analyzed a large set of
cosmological $N$-body simulations to test how accurately this
formalism predicts the true power spectrum in redshift space and how
many terms are needed to achieve a given precision.  We compared to the
RSD power spectrum as a function of wavevector $k$ and cosine of the
angle between the Fourier mode and line of sight $\mu$, as well as to
the lowest Legendre moments, monopole, quadrupole, and hexadecapole.
These comparisons revealed that the expansion is accurate within a few
percent up to $k\simeq 0.2 \hmpci$ at $z=0$ and $k\simeq 0.3 \hmpci$
at $z=2$, if the corrections up to the 6th order are taken into
account.  All expansions break down at higher $k$, where the expansion
parameter becomes larger than unity.

The expansion can be resummed into power suppression factors known as
the Finger-of-God (FoG) kernels. Our expansion formula suggests that
one needs three independent FoG kernels, each of which is multiplying
the lowest 3 terms in the expansion, which are the 3 terms that
contain linear order contributions (Kaiser formula).  We have found that the
3 velocity dispersions differ in their values. We also found that the
shape of the FoG kernels differ and are a function of scale, redshift
and angle. There is no single FoG kernel that would fit the simulation
data.  This FoG model has validity comparable or better than the
highest (6th) order summations we have tried.  Our FoG model predicts
the monopole power spectrum with a few percent accuracy up to $k\simeq
0.4 \hmpci$ at $z=0$ and is accurate for $k<1 \hmpci$ at $z=0.5$ and 1.

The difficulty of obtaining convergence or a perfect FoG model on
small scales in Legendre moments expansion do not exist in the
alternative expansion in powers of $\mu^{2j}$. This expansion has the
advantage of containing a finite number of terms for a given value of
$j$, and hence no infinite sums are needed, there is always
convergence for any finite $j$. In contrast every Legendre moment
receives infinite number of terms. We have used the angular decomposition of
\cite{Seljak:2011} to derive the individual terms for the lowest
values of $j$. We compared their amplitudes as a function of scale,
showing that for $\mu^2$ and $\mu^4$ the nonlinear FoG terms dominate
the Kaiser terms for $k>0.2\hmpci$ ($z=0$).  Similarly, higher orders
($\mu^6$ etc), which do not receive any contributions at the linear
order, are nearly zero for $k \ll 0.2\hmpci$, but grow very rapidly on
smaller scales. Because the higher order terms of $P_{LL'}$ also have 
higher order powers of $\mu^2$, one can in principle separate them from 
lower order terms, put them all into terms that we do not care about, 
and we can marginalize over. 
Note that the expansion in powers of $\mu^2$ has the
disadvantage that the coefficients extracted from observations will be
correlated, unlike Legendre moments, which are orthogonal, at least on
scales small compared to the size of the survey. In practice this
means that some priors will have to be assumed for the terms with
higher powers of $\mu^2$, at least at high $k$, otherwise they will
contaminate the determination of $\mu^0$, $\mu^2$ and $\mu^4$
terms. In this sense we cannot escape the high order FoG
complications. However, it is not clear there is any useful
information remaining at $k>0.2\hmpci$ ($z=0$) due to the nonlinear
evolution anyways. This will be explored further with perturbation
theory in paper III of this series \cite{Vlah:2012}.

We also compared our approach based on the power spectrum of
mass-weighted velocity moments to the one based on the power spectrum
of volume-weighted velocity moments.  We found that the power
spectrum based on the mass-weighted velocity moments converges
slightly faster, although the difference is not so significant. 
The fact that there is no natural way to expand RSD into
volume weighted quantities, and that using volume weighted quantities
does not improve the convergence, implies that there is no good reason
to work with volume weighted quantities in RSD studies.

This advantage of mass weighting quantities becomes crucial when the
same formalism is applied to galaxies and halos, where volume weighted
moments cannot be easily defined.  It is difficult or impossible to
measure the velocity power spectrum from the sparse density field and
the results depend on the smoothing (see e.g.  \cite{Tang:2011} for an
attempt and failure to measure it).
In contrast to this, the analysis presented in this paper can be
naturally extended to the clustering analysis of halos and galaxies,
since density weighted moments are not affected by sparseness of the
sample.  Recently it was shown by \cite{Okumura:2011} (but see also
\cite{Reid:2011}) that the RSD parameters reconstracted from
redshift-space distortions of dark matter halos have strong halo-mass
and scale dependence even on large scales ($k<0.1\hmpci$).  This can
be a serious problem when one wants to constrain dark energy or
modified gravity theories.  This complicated scale dependence is
likely to be a consequence of the number density weighting of velocity
moments, which differs from the mass weighting of dark matter if
galaxies or halos have a different spatial distribution.  The scale
dependence will arise even for a linear bias model, except on very
large scales, where galaxy overdensity $\delta_g \ll 1$
\cite{Desjacques:2010a, Seljak:2011}.  The formalism presented in this paper can be used
to investigate this scale dependence and we plan to present this
analysis in paper IV of this series.

\acknowledgments We would like to thank Nico Hamaus, Zvonimir Vlah and
Tobias Baldauf for help and useful discussions.  This research was
supported by the DOE, and the Swiss National Foundation under contract
200021-116696/1 and Republic of Korea WCU grant R32-10130.

\appendix

\section{Derivatives of FoG kernels from higher-order $P_{LL'}$'s} \label{sec:derivative}
In this appendix, we present derivatives of the FoG kernels: $G_{00}$, $G_{0 1}$ and $G_{11}$ for
the auto power spectrum of the redshift-space density field,
\begin{eqnarray}
	\left. G^{(1)}_{00}(x_{00})\right |_{x_{0 0}=0} &=& -\frac{1}{\sigma_{00}^2} \frac{P_{0 2}}{P_{00}}, \\
	\left. G^{(1)}_{0 1}(x_{0 1})      \right |_{x_{0 1}=0} &=& -\frac{1}{\sigma_{0 1}^2P_{0 1}} \left[ \frac{1}{6}P_{0 3} + \frac{1}{2}P_{21} \right], \\	
	\left. G^{(1)}_{11}(x_{11}) \right |_{x_{11}=0} &=& -\frac{1}{3\sigma_{11}^2} \frac{P_{13}}{P_{11}}, \\
	\left. G^{(2)}_{00}(x_{00})\right |_{x_{0 0}=0} &=&  \frac{1}{\sigma_{00}^4P_{00}} \left[ \frac{1}{6}P_{0 4} + \frac{1}{2}P_{22} \right], \label{eq:G_00_2} \\
	\left. G^{(2)}_{0 1}(x_{0 1})      \right |_{x_{0 1}=0} &=&  \frac{1}{\sigma_{0 1}^4P_{0 1}} \left[ \frac{1}{60}P_{0 5} + \frac{1}{12}P_{41} + \frac{1}{6}P_{23} \right], \\
	\left. G^{(2)}_{11}(x_{11}) \right |_{x_{11}=0} &=&  \frac{1}{\sigma_{11}^4P_{11}} \left[ \frac{1}{30}P_{15} + \frac{1}{16}P_{33} \right], \label{eq:G_11_2} \\
	\left. G^{(3)}_{00}(x_{00})\right |_{x_{0 0}=0} &=& -\frac{1}{\sigma_{00}^6P_{00}} \left[ \frac{1}{60}P_{0 6} + \frac{1}{4}P_{24} \right], \\	
	\left. G^{(3)}_{0 1}(x_{0 1})      \right |_{x_{0 1}=0} &=& -\frac{1}{\sigma_{0 1}^6P_{0 1}} \left[ \frac{1}{840}P_{0 7} 
		+ \frac{1}{120}P_{61} + \frac{1}{40}P_{25}+ \frac{1}{24}P_{43}  \right], \\
	\left. G^{(3)}_{11}(x_{11}) \right |_{x_{11}=0} &=& -\frac{1}{\sigma_{11}^6P_{11}} \left[ \frac{1}{420}P_{17} + \frac{1}{60}P_{35} \right], \\
	\left. G^{(4)}_{00}(x_{00})\right |_{x_{0 0}=0} &=&  \frac{1}{\sigma_{00}^8P_{00}} \left[ \frac{1}{840}P_{0 8} 
		+ \frac{1}{30}P_{26} + \frac{1}{24}P_{44} \right], \\
	\left. G^{(4)}_{0 1}(x_{0 1})      \right |_{x_{0 1}=0} &=&  \frac{1}{\sigma_{11}^8P_{0 1}} \left[ \frac{1}{15120}P_{0 9} + \frac{1}{1680}P_{81} 
		+ \frac{1}{420}P_{27}+ \frac{1}{180}P_{63} + \frac{1}{120}P_{45}  \right], \\
	\left. G^{(4)}_{11}(x_{11}) \right |_{x_{11}=0} &=&  \frac{1}{\sigma_{11}^8P_{11}} \left[ \frac{1}{7560}P_{19} + \frac{1}{630}P_{37} + \frac{1}{600}P_{55} \right].
\end{eqnarray}
Note that these equations are general and not dependent of the ansatz
of the kernel (equation (\ref{eq:fog})).
As discussed in section \ref{sec:fog_resum}, the first derivatives of
$G_{LL'}$ give the definitions of the velocity dispersions
$\sigma_{LL'}$.
The power low index parameters of our FoG model, $\alpha_{LL'}$, are obtained by substituting 
equations (\ref{eq:G_00_2}) -- (\ref{eq:G_11_2}) into equation (\ref{eq:alpha_def}).

%\bibliography{rsd}
\bibliography{ms.bbl}

\begin{thebibliography}{10}
\providecommand*{\bibinfo}[2]{#2}
\providecommand*{\eprint}[1]{#1}
\providecommand*{\url}[1]{#1}
\bibitem{Peebles:1980}
\bibinfo{author}{P.~J.~E. {Peebles}}, \bibinfo{title}{\emph{{The large-scale
  structure of the universe}}} (\bibinfo{year}{1980}).
\bibitem{Eisenstein:2005}
\bibinfo{author}{D.~J. {Eisenstein}}, \bibinfo{author}{I.~{Zehavi}},
  \bibinfo{author}{D.~W. {Hogg}}, \bibinfo{author}{R.~{Scoccimarro}},
  \bibinfo{author}{M.~R. {Blanton}}, \bibinfo{author}{R.~C. {Nichol}},
  \bibinfo{author}{R.~{Scranton}}, \bibinfo{author}{H.~{Seo}},
  \bibinfo{author}{M.~{Tegmark}}, \bibinfo{author}{Z.~{Zheng}}, \emph{et~al.},
  \bibinfo{journal}{\apj} \bibinfo{volume}{\textbf{633}}, \bibinfo{pages}{560}
  (\bibinfo{date}{Nov. 2005}), \eprint{arXiv:astro-ph/0501171}.
\bibitem{Cole:2005}
\bibinfo{author}{S.~{Cole}}, \bibinfo{author}{W.~J. {Percival}},
  \bibinfo{author}{J.~A. {Peacock}}, \bibinfo{author}{P.~{Norberg}},
  \bibinfo{author}{C.~M. {Baugh}}, \bibinfo{author}{C.~S. {Frenk}},
  \bibinfo{author}{I.~{Baldry}}, \bibinfo{author}{J.~{Bland-Hawthorn}},
  \bibinfo{author}{T.~{Bridges}}, \bibinfo{author}{R.~{Cannon}}, \emph{et~al.},
  \bibinfo{journal}{\mnras} \bibinfo{volume}{\textbf{362}},
  \bibinfo{pages}{505} (\bibinfo{date}{Sep. 2005}),
  \eprint{arXiv:astro-ph/0501174}.
\bibitem{Jackson:1972}
\bibinfo{author}{J.~C. {Jackson}}, \bibinfo{journal}{\mnras}
  \bibinfo{volume}{\textbf{156}}, \bibinfo{pages}{1P} (\bibinfo{date}{1972}).
\bibitem{Kaiser:1987}
\bibinfo{author}{N.~{Kaiser}}, \bibinfo{journal}{\mnras}
  \bibinfo{volume}{\textbf{227}}, \bibinfo{pages}{1} (\bibinfo{date}{Jul.
  1987}).
\bibitem{Hamilton:1998}
\bibinfo{author}{A.~J.~S. {Hamilton}}, in \bibinfo{editors}{{D.~Hamilton}},
  ed., \emph{The Evolving Universe} (\bibinfo{date}{1998}),
  \bibinfo{volume}{vol. 231 of \emph{Astrophysics and Space Science Library}},
  \bibinfo{pages}{pp. 185--+}, \eprint{arXiv:astro-ph/9708102}.
\bibitem{Linder:2005}
\bibinfo{author}{E.~V. {Linder}}, \bibinfo{journal}{\prd}
  \bibinfo{volume}{\textbf{72}}(4), \bibinfo{pages}{043529}
  (\bibinfo{date}{Aug. 2005}), \eprint{arXiv:astro-ph/0507263}.
\bibitem{Jain:2008}
\bibinfo{author}{B.~{Jain}} and \bibinfo{author}{P.~{Zhang}},
  \bibinfo{journal}{\prd} \bibinfo{volume}{\textbf{78}}(6),
  \bibinfo{pages}{063503} (\bibinfo{date}{Sep. 2008}), \eprint{0709.2375}.
\bibitem{Song:2009}
\bibinfo{author}{Y.~{Song}} and \bibinfo{author}{K.~{Koyama}},
  \bibinfo{journal}{\jcap} \bibinfo{volume}{\textbf{1}}, \bibinfo{pages}{48}
  (\bibinfo{date}{Jan. 2009}), \eprint{0802.3897}.
\bibitem{Alcock:1979}
\bibinfo{author}{C.~{Alcock}} and \bibinfo{author}{B.~{Paczynski}},
  \bibinfo{journal}{\nat} \bibinfo{volume}{\textbf{281}}, \bibinfo{pages}{358}
  (\bibinfo{date}{Oct. 1979}).
\bibitem{Matsubara:1996}
\bibinfo{author}{T.~{Matsubara}} and \bibinfo{author}{Y.~{Suto}},
  \bibinfo{journal}{\apjl} \bibinfo{volume}{\textbf{470}}, \bibinfo{pages}{L1+}
  (\bibinfo{date}{Oct. 1996}), \eprint{arXiv:astro-ph/9604142}.
\bibitem{Ballinger:1996}
\bibinfo{author}{W.~E. {Ballinger}}, \bibinfo{author}{J.~A. {Peacock}}, and
  \bibinfo{author}{A.~F. {Heavens}}, \bibinfo{journal}{\mnras}
  \bibinfo{volume}{\textbf{282}}, \bibinfo{pages}{877} (\bibinfo{date}{Oct.
  1996}), \eprint{arXiv:astro-ph/9605017}.
\bibitem{Peacock:2001}
\bibinfo{author}{J.~A. {Peacock}}, \bibinfo{author}{S.~{Cole}},
  \bibinfo{author}{P.~{Norberg}}, \bibinfo{author}{C.~M. {Baugh}},
  \bibinfo{author}{J.~{Bland-Hawthorn}}, \bibinfo{author}{T.~{Bridges}},
  \bibinfo{author}{R.~D. {Cannon}}, \bibinfo{author}{M.~{Colless}},
  \bibinfo{author}{C.~{Collins}}, \bibinfo{author}{W.~{Couch}}, \emph{et~al.},
  \bibinfo{journal}{\nat} \bibinfo{volume}{\textbf{410}}, \bibinfo{pages}{169}
  (\bibinfo{date}{Mar. 2001}), \eprint{arXiv:astro-ph/0103143}.
\bibitem{Zehavi:2002}
\bibinfo{author}{I.~{Zehavi}}, \bibinfo{author}{M.~R. {Blanton}},
  \bibinfo{author}{J.~A. {Frieman}}, \bibinfo{author}{D.~H. {Weinberg}},
  \bibinfo{author}{H.~J. {Mo}}, \bibinfo{author}{M.~A. {Strauss}},
  \bibinfo{author}{S.~F. {Anderson}}, \bibinfo{author}{J.~{Annis}},
  \bibinfo{author}{N.~A. {Bahcall}}, \bibinfo{author}{M.~{Bernardi}},
  \emph{et~al.}, \bibinfo{journal}{\apj} \bibinfo{volume}{\textbf{571}},
  \bibinfo{pages}{172} (\bibinfo{date}{May 2002}),
  \eprint{arXiv:astro-ph/0106476}.
\bibitem{Hawkins:2003}
\bibinfo{author}{E.~{Hawkins}}, \bibinfo{author}{S.~{Maddox}},
  \bibinfo{author}{S.~{Cole}}, \bibinfo{author}{O.~{Lahav}},
  \bibinfo{author}{D.~S. {Madgwick}}, \bibinfo{author}{P.~{Norberg}},
  \bibinfo{author}{J.~A. {Peacock}}, \bibinfo{author}{I.~K. {Baldry}},
  \bibinfo{author}{C.~M. {Baugh}}, \bibinfo{author}{J.~{Bland-Hawthorn}},
  \emph{et~al.}, \bibinfo{journal}{\mnras} \bibinfo{volume}{\textbf{346}},
  \bibinfo{pages}{78} (\bibinfo{date}{Nov. 2003}),
  \eprint{arXiv:astro-ph/0212375}.
\bibitem{Tegmark:2004}
\bibinfo{author}{M.~{Tegmark}}, \bibinfo{author}{M.~A. {Strauss}},
  \bibinfo{author}{M.~R. {Blanton}}, \bibinfo{author}{K.~{Abazajian}},
  \bibinfo{author}{S.~{Dodelson}}, \bibinfo{author}{H.~{Sandvik}},
  \bibinfo{author}{X.~{Wang}}, \bibinfo{author}{D.~H. {Weinberg}},
  \bibinfo{author}{I.~{Zehavi}}, \bibinfo{author}{N.~A. {Bahcall}},
  \emph{et~al.}, \bibinfo{journal}{\prd} \bibinfo{volume}{\textbf{69}}(10),
  \bibinfo{pages}{103501} (\bibinfo{date}{May 2004}),
  \eprint{arXiv:astro-ph/0310723}.
\bibitem{Tegmark:2006}
\bibinfo{author}{M.~{Tegmark}}, \bibinfo{author}{D.~J. {Eisenstein}},
  \bibinfo{author}{M.~A. {Strauss}}, \bibinfo{author}{D.~H. {Weinberg}},
  \bibinfo{author}{M.~R. {Blanton}}, \bibinfo{author}{J.~A. {Frieman}},
  \bibinfo{author}{M.~{Fukugita}}, \bibinfo{author}{J.~E. {Gunn}},
  \bibinfo{author}{A.~J.~S. {Hamilton}}, \bibinfo{author}{G.~R. {Knapp}},
  \emph{et~al.}, \bibinfo{journal}{\prd} \bibinfo{volume}{\textbf{74}}(12),
  \bibinfo{pages}{123507} (\bibinfo{date}{Dec. 2006}),
  \eprint{arXiv:astro-ph/0608632}.
\bibitem{Ross:2007}
\bibinfo{author}{N.~P. {Ross}}, \bibinfo{author}{J.~{da {\^A}ngela}},
  \bibinfo{author}{T.~{Shanks}}, \bibinfo{author}{D.~A. {Wake}},
  \bibinfo{author}{R.~D. {Cannon}}, \bibinfo{author}{A.~C. {Edge}},
  \bibinfo{author}{R.~C. {Nichol}}, \bibinfo{author}{P.~J. {Outram}},
  \bibinfo{author}{M.~{Colless}}, \bibinfo{author}{W.~J. {Couch}},
  \emph{et~al.}, \bibinfo{journal}{\mnras} \bibinfo{volume}{\textbf{381}},
  \bibinfo{pages}{573} (\bibinfo{date}{Oct. 2007}),
  \eprint{arXiv:astro-ph/0612400}.
\bibitem{Guzzo:2008}
\bibinfo{author}{L.~{Guzzo}}, \bibinfo{author}{M.~{Pierleoni}},
  \bibinfo{author}{B.~{Meneux}}, \bibinfo{author}{E.~{Branchini}},
  \bibinfo{author}{O.~{Le F{\`e}vre}}, \bibinfo{author}{C.~{Marinoni}},
  \bibinfo{author}{B.~{Garilli}}, \bibinfo{author}{J.~{Blaizot}},
  \bibinfo{author}{G.~{De Lucia}}, \bibinfo{author}{A.~{Pollo}}, \emph{et~al.},
  \bibinfo{journal}{\nat} \bibinfo{volume}{\textbf{451}}, \bibinfo{pages}{541}
  (\bibinfo{date}{Jan. 2008}), \eprint{0802.1944}.
\bibitem{Okumura:2008}
\bibinfo{author}{T.~{Okumura}}, \bibinfo{author}{T.~{Matsubara}},
  \bibinfo{author}{D.~J. {Eisenstein}}, \bibinfo{author}{I.~{Kayo}},
  \bibinfo{author}{C.~{Hikage}}, \bibinfo{author}{A.~S. {Szalay}}, and
  \bibinfo{author}{D.~P. {Schneider}}, \bibinfo{journal}{\apj}
  \bibinfo{volume}{\textbf{676}}, \bibinfo{pages}{889} (\bibinfo{date}{Apr.
  2008}), \eprint{0711.3640}.
\bibitem{Cabre:2009}
\bibinfo{author}{A.~{Cabr{\'e}}} and \bibinfo{author}{E.~{Gazta{\~n}aga}},
  \bibinfo{journal}{\mnras} \bibinfo{volume}{\textbf{393}},
  \bibinfo{pages}{1183} (\bibinfo{date}{Mar. 2009}), \eprint{0807.2460}.
\bibitem{Blake:2011}
\bibinfo{author}{C.~{Blake}}, \bibinfo{author}{S.~{Brough}},
  \bibinfo{author}{M.~{Colless}}, \bibinfo{author}{C.~{Contreras}},
  \bibinfo{author}{W.~{Couch}}, \bibinfo{author}{S.~{Croom}},
  \bibinfo{author}{T.~{Davis}}, \bibinfo{author}{M.~J. {Drinkwater}},
  \bibinfo{author}{K.~{Forster}}, \bibinfo{author}{D.~{Gilbank}},
  \emph{et~al.}, \bibinfo{journal}{\mnras} \bibinfo{pages}{pp. 834--+}
  (\bibinfo{date}{Jun. 2011}), \eprint{1104.2948}.
\bibitem{Tinker:2006}
\bibinfo{author}{J.~L. {Tinker}}, \bibinfo{author}{D.~H. {Weinberg}}, and
  \bibinfo{author}{Z.~{Zheng}}, \bibinfo{journal}{\mnras}
  \bibinfo{volume}{\textbf{368}}, \bibinfo{pages}{85} (\bibinfo{date}{May
  2006}), \eprint{arXiv:astro-ph/0501029}.
\bibitem{Okumura:2011}
\bibinfo{author}{T.~{Okumura}} and \bibinfo{author}{Y.~P. {Jing}},
  \bibinfo{journal}{\apj} \bibinfo{volume}{\textbf{726}}, \bibinfo{pages}{5}
  (\bibinfo{date}{Jan. 2011}), \eprint{1004.3548}.
\bibitem{Jennings:2011}
\bibinfo{author}{E.~{Jennings}}, \bibinfo{author}{C.~M. {Baugh}}, and
  \bibinfo{author}{S.~{Pascoli}}, \bibinfo{journal}{\mnras}
  \bibinfo{volume}{\textbf{410}}, \bibinfo{pages}{2081} (\bibinfo{date}{Jan.
  2011}), \eprint{1003.4282}.
\bibitem{Kwan:2011}
\bibinfo{author}{J.~{Kwan}}, \bibinfo{author}{G.~F. {Lewis}}, and
  \bibinfo{author}{E.~V. {Linder}}, \bibinfo{journal}{ArXiv e-prints}
  (\bibinfo{date}{May 2011}), \eprint{1105.1194}.
\bibitem{Hill:2004}
\bibinfo{author}{G.~J. {Hill}}, \bibinfo{author}{K.~{Gebhardt}},
  \bibinfo{author}{E.~{Komatsu}}, and \bibinfo{author}{P.~J. {MacQueen}}, in
  \bibinfo{editors}{{R.~E.~Allen, D.~V.~Nanopoulos, \& C.~N.~Pope}}, ed.,
  \emph{The New Cosmology: Conference on Strings and Cosmology}
  (\bibinfo{date}{Dec. 2004}), \bibinfo{volume}{vol. 743 of \emph{American
  Institute of Physics Conference Series}}, \bibinfo{pages}{pp. 224--233}.
\bibitem{Glazebrook:2007}
\bibinfo{author}{K.~{Glazebrook}}, \bibinfo{author}{C.~{Blake}},
  \bibinfo{author}{W.~{Couch}}, \bibinfo{author}{D.~{Forbes}},
  \bibinfo{author}{M.~{Drinkwater}}, \bibinfo{author}{R.~{Jurek}},
  \bibinfo{author}{K.~{Pimbblet}}, \bibinfo{author}{B.~{Madore}},
  \bibinfo{author}{C.~{Martin}}, \bibinfo{author}{T.~{Small}}, \emph{et~al.},
  \bibinfo{journal}{ArXiv Astrophysics e-prints}  (\bibinfo{date}{Jan. 2007}),
  \eprint{arXiv:astro-ph/0701876}.
\bibitem{Schlegel:2009}
\bibinfo{author}{D.~{Schlegel}}, \bibinfo{author}{M.~{White}}, and
  \bibinfo{author}{D.~{Eisenstein}}, in \emph{astro2010: The Astronomy and
  Astrophysics Decadal Survey} (\bibinfo{date}{2009}), \bibinfo{volume}{vol.
  2010 of \emph{ArXiv Astrophysics e-prints}}, \bibinfo{pages}{pp. 314--+},
  \eprint{0902.4680}.
\bibitem{Sumiyoshi:2009}
\bibinfo{author}{M.~{Sumiyoshi}}, \bibinfo{author}{T.~{Totani}},
  \bibinfo{author}{S.~{Oshige}}, \bibinfo{author}{K.~{Glazebrook}},
  \bibinfo{author}{M.~{Akiyama}}, \bibinfo{author}{T.~{Morokuma}},
  \bibinfo{author}{K.~{Motohara}}, \bibinfo{author}{K.~{Shimasaku}},
  \bibinfo{author}{M.~{Hayashi}}, \bibinfo{author}{M.~{Yoshida}},
  \emph{et~al.}, \bibinfo{journal}{ArXiv e-prints}  (\bibinfo{date}{Feb.
  2009}), \eprint{0902.2064}.
\bibitem{Schlegel:2009a}
\bibinfo{author}{D.~J. {Schlegel}}, \bibinfo{author}{C.~{Bebek}},
  \bibinfo{author}{H.~{Heetderks}}, \bibinfo{author}{S.~{Ho}},
  \bibinfo{author}{M.~{Lampton}}, \bibinfo{author}{M.~{Levi}},
  \bibinfo{author}{N.~{Mostek}}, \bibinfo{author}{N.~{Padmanabhan}},
  \bibinfo{author}{S.~{Perlmutter}}, \bibinfo{author}{N.~{Roe}}, \emph{et~al.},
  \bibinfo{journal}{ArXiv e-prints}  (\bibinfo{date}{Apr. 2009}),
  \eprint{0904.0468}.
\bibitem{Scoccimarro:2004}
\bibinfo{author}{R.~{Scoccimarro}}, \bibinfo{journal}{\prd}
  \bibinfo{volume}{\textbf{70}}(8), \bibinfo{pages}{083007}
  (\bibinfo{date}{Oct. 2004}), \eprint{arXiv:astro-ph/0407214}.
\bibitem{Bernardeau:2002}
\bibinfo{author}{F.~{Bernardeau}}, \bibinfo{author}{S.~{Colombi}},
  \bibinfo{author}{E.~{Gazta{\~n}aga}}, and \bibinfo{author}{R.~{Scoccimarro}},
  \bibinfo{journal}{\physrep} \bibinfo{volume}{\textbf{367}},
  \bibinfo{pages}{1} (\bibinfo{date}{Sep. 2002}),
  \eprint{arXiv:astro-ph/0112551}.
\bibitem{Crocce:2006}
\bibinfo{author}{M.~{Crocce}} and \bibinfo{author}{R.~{Scoccimarro}},
  \bibinfo{journal}{\prd} \bibinfo{volume}{\textbf{73}}(6),
  \bibinfo{pages}{063519} (\bibinfo{date}{Mar. 2006}),
  \eprint{arXiv:astro-ph/0509418}.
\bibitem{Crocce:2006b}
\bibinfo{author}{M.~{Crocce}} and \bibinfo{author}{R.~{Scoccimarro}},
  \bibinfo{journal}{\prd} \bibinfo{volume}{\textbf{73}}(6),
  \bibinfo{pages}{063520} (\bibinfo{date}{Mar. 2006}),
  \eprint{arXiv:astro-ph/0509419}.
\bibitem{Matarrese:2007}
\bibinfo{author}{S.~{Matarrese}} and \bibinfo{author}{M.~{Pietroni}},
  \bibinfo{journal}{\jcap} \bibinfo{volume}{\textbf{6}}, \bibinfo{pages}{26}
  (\bibinfo{date}{Jun. 2007}), \eprint{arXiv:astro-ph/0703563}.
\bibitem{McDonald:2007}
\bibinfo{author}{P.~{McDonald}}, \bibinfo{journal}{\prd}
  \bibinfo{volume}{\textbf{75}}(4), \bibinfo{pages}{043514}
  (\bibinfo{date}{Feb. 2007}), \eprint{arXiv:astro-ph/0606028}.
\bibitem{Valageas:2007}
\bibinfo{author}{P.~{Valageas}}, \bibinfo{journal}{\aap}
  \bibinfo{volume}{\textbf{465}}, \bibinfo{pages}{725} (\bibinfo{date}{Apr.
  2007}), \eprint{arXiv:astro-ph/0611849}.
\bibitem{Taruya:2008}
\bibinfo{author}{A.~{Taruya}} and \bibinfo{author}{T.~{Hiramatsu}},
  \bibinfo{journal}{\apj} \bibinfo{volume}{\textbf{674}}, \bibinfo{pages}{617}
  (\bibinfo{date}{Feb. 2008}), \eprint{0708.1367}.
\bibitem{Nishimichi:2009}
\bibinfo{author}{T.~{Nishimichi}}, \bibinfo{author}{A.~{Shirata}},
  \bibinfo{author}{A.~{Taruya}}, \bibinfo{author}{K.~{Yahata}},
  \bibinfo{author}{S.~{Saito}}, \bibinfo{author}{Y.~{Suto}},
  \bibinfo{author}{R.~{Takahashi}}, \bibinfo{author}{N.~{Yoshida}},
  \bibinfo{author}{T.~{Matsubara}}, \bibinfo{author}{N.~{Sugiyama}},
  \emph{et~al.}, \bibinfo{journal}{\pasj} \bibinfo{volume}{\textbf{61}},
  \bibinfo{pages}{321} (\bibinfo{date}{Feb. 2009}), \eprint{0810.0813}.
\bibitem{Carlson:2009}
\bibinfo{author}{J.~{Carlson}}, \bibinfo{author}{M.~{White}}, and
  \bibinfo{author}{N.~{Padmanabhan}}, \bibinfo{journal}{\prd}
  \bibinfo{volume}{\textbf{80}}(4), \bibinfo{pages}{043531}
  (\bibinfo{date}{Aug. 2009}), \eprint{0905.0479}.
\bibitem{Heavens:1998}
\bibinfo{author}{A.~F. {Heavens}}, \bibinfo{author}{S.~{Matarrese}}, and
  \bibinfo{author}{L.~{Verde}}, \bibinfo{journal}{\mnras}
  \bibinfo{volume}{\textbf{301}}, \bibinfo{pages}{797} (\bibinfo{date}{Dec.
  1998}), \eprint{arXiv:astro-ph/9808016}.
\bibitem{Scoccimarro:1999}
\bibinfo{author}{R.~{Scoccimarro}}, \bibinfo{author}{H.~M.~P. {Couchman}}, and
  \bibinfo{author}{J.~A. {Frieman}}, \bibinfo{journal}{\apj}
  \bibinfo{volume}{\textbf{517}}, \bibinfo{pages}{531} (\bibinfo{date}{Jun.
  1999}), \eprint{arXiv:astro-ph/9808305}.
\bibitem{Bharadwaj:2001}
\bibinfo{author}{S.~{Bharadwaj}}, \bibinfo{journal}{\mnras}
  \bibinfo{volume}{\textbf{327}}, \bibinfo{pages}{577} (\bibinfo{date}{Oct.
  2001}), \eprint{arXiv:astro-ph/0105320}.
\bibitem{Pandey:2005}
\bibinfo{author}{B.~{Pandey}} and \bibinfo{author}{S.~{Bharadwaj}},
  \bibinfo{journal}{\mnras} \bibinfo{volume}{\textbf{358}},
  \bibinfo{pages}{939} (\bibinfo{date}{Apr. 2005}),
  \eprint{arXiv:astro-ph/0403670}.
\bibitem{Matsubara:2008}
\bibinfo{author}{T.~{Matsubara}}, \bibinfo{journal}{\prd}
  \bibinfo{volume}{\textbf{77}}(6), \bibinfo{pages}{063530}
  (\bibinfo{date}{Mar. 2008}), \eprint{0711.2521}.
\bibitem{Taruya:2010}
\bibinfo{author}{A.~{Taruya}}, \bibinfo{author}{T.~{Nishimichi}}, and
  \bibinfo{author}{S.~{Saito}}, \bibinfo{journal}{\prd}
  \bibinfo{volume}{\textbf{82}}(6), \bibinfo{pages}{063522}
  (\bibinfo{date}{Sep. 2010}), \eprint{1006.0699}.
\bibitem{Valageas:2011}
\bibinfo{author}{P.~{Valageas}}, \bibinfo{journal}{\aap}
  \bibinfo{volume}{\textbf{526}}, \bibinfo{pages}{A67+} (\bibinfo{date}{Feb.
  2011}), \eprint{1009.0106}.
\bibitem{Seljak:2011}
\bibinfo{author}{U.~{Seljak}} and \bibinfo{author}{P.~{McDonald}},
  \bibinfo{journal}{\jcap} \bibinfo{volume}{\textbf{11}}, \bibinfo{pages}{39}
  (\bibinfo{date}{Nov. 2011}), \eprint{1109.1888}.
\bibitem{McDonald:2011}
\bibinfo{author}{P.~{McDonald}}, \bibinfo{journal}{\jcap}
  \bibinfo{volume}{\textbf{4}}, \bibinfo{pages}{32} (\bibinfo{date}{Apr.
  2011}), \eprint{0910.1002}.
\bibitem{Desjacques:2009}
\bibinfo{author}{V.~{Desjacques}}, \bibinfo{author}{U.~{Seljak}}, and
  \bibinfo{author}{I.~T. {Iliev}}, \bibinfo{journal}{\mnras}
  \bibinfo{volume}{\textbf{396}}, \bibinfo{pages}{85} (\bibinfo{date}{Jun.
  2009}), \eprint{0811.2748}.
\bibitem{Seljak:1996}
\bibinfo{author}{U.~{Seljak}} and \bibinfo{author}{M.~{Zaldarriaga}},
  \bibinfo{journal}{\apj} \bibinfo{volume}{\textbf{469}}, \bibinfo{pages}{437}
  (\bibinfo{date}{Oct. 1996}), \eprint{arXiv:astro-ph/9603033}.
\bibitem{Komatsu:2009}
\bibinfo{author}{E.~{Komatsu}}, \bibinfo{author}{J.~{Dunkley}},
  \bibinfo{author}{M.~R. {Nolta}}, \bibinfo{author}{C.~L. {Bennett}},
  \bibinfo{author}{B.~{Gold}}, \bibinfo{author}{G.~{Hinshaw}},
  \bibinfo{author}{N.~{Jarosik}}, \bibinfo{author}{D.~{Larson}},
  \bibinfo{author}{M.~{Limon}}, \bibinfo{author}{L.~{Page}}, \emph{et~al.},
  \bibinfo{journal}{\apjs} \bibinfo{volume}{\textbf{180}}, \bibinfo{pages}{330}
  (\bibinfo{date}{Feb. 2009}), \eprint{0803.0547}.
\bibitem{McDonald:2009}
\bibinfo{author}{P.~{McDonald}} and \bibinfo{author}{U.~{Seljak}},
  \bibinfo{journal}{\jcap} \bibinfo{volume}{\textbf{10}}, \bibinfo{pages}{7}
  (\bibinfo{date}{Oct. 2009}), \eprint{0810.0323}.
\bibitem{Yamamoto:2005}
\bibinfo{author}{K.~{Yamamoto}}, \bibinfo{author}{B.~A. {Bassett}}, and
  \bibinfo{author}{H.~{Nishioka}}, \bibinfo{journal}{Physical Review Letters}
  \bibinfo{volume}{\textbf{94}}(5), \bibinfo{pages}{051301}
  (\bibinfo{date}{Feb. 2005}), \eprint{arXiv:astro-ph/0409207}.
\bibitem{Tocchini-Valentini:2011}
\bibinfo{author}{D.~{Tocchini-Valentini}}, \bibinfo{author}{M.~{Barnard}},
  \bibinfo{author}{C.~L. {Bennett}}, and \bibinfo{author}{A.~S. {Szalay}},
  \bibinfo{journal}{ArXiv e-prints}  (\bibinfo{date}{Jan. 2011}),
  \eprint{1101.2608}.
\bibitem{Taruya:2011}
\bibinfo{author}{A.~{Taruya}}, \bibinfo{author}{S.~{Saito}}, and
  \bibinfo{author}{T.~{Nishimichi}}, \bibinfo{journal}{\prd}
  \bibinfo{volume}{\textbf{83}}(10), \bibinfo{pages}{103527}
  (\bibinfo{date}{May 2011}), \eprint{1101.4723}.
\bibitem{Cole:1994}
\bibinfo{author}{S.~{Cole}}, \bibinfo{author}{K.~B. {Fisher}}, and
  \bibinfo{author}{D.~H. {Weinberg}}, \bibinfo{journal}{\mnras}
  \bibinfo{volume}{\textbf{267}}, \bibinfo{pages}{785} (\bibinfo{date}{Apr.
  1994}), \eprint{arXiv:astro-ph/9308003}.
\bibitem{Eisenstein:1998}
\bibinfo{author}{D.~J. {Eisenstein}} and \bibinfo{author}{W.~{Hu}},
  \bibinfo{journal}{\apj} \bibinfo{volume}{\textbf{496}}, \bibinfo{pages}{605}
  (\bibinfo{date}{Mar. 1998}), \eprint{arXiv:astro-ph/9709112}.
\bibitem{Peacock:1994}
\bibinfo{author}{J.~A. {Peacock}} and \bibinfo{author}{S.~J. {Dodds}},
  \bibinfo{journal}{\mnras} \bibinfo{volume}{\textbf{267}},
  \bibinfo{pages}{1020} (\bibinfo{date}{Apr. 1994}),
  \eprint{arXiv:astro-ph/9311057}.
\bibitem{Park:1994}
\bibinfo{author}{C.~{Park}}, \bibinfo{author}{M.~S. {Vogeley}},
  \bibinfo{author}{M.~J. {Geller}}, and \bibinfo{author}{J.~P. {Huchra}},
  \bibinfo{journal}{\apj} \bibinfo{volume}{\textbf{431}}, \bibinfo{pages}{569}
  (\bibinfo{date}{Aug. 1994}).
\bibitem{Percival:2009}
\bibinfo{author}{W.~J. {Percival}} and \bibinfo{author}{M.~{White}},
  \bibinfo{journal}{\mnras} \bibinfo{volume}{\textbf{393}},
  \bibinfo{pages}{297} (\bibinfo{date}{Feb. 2009}), \eprint{0808.0003}.
\bibitem{Tang:2011}
\bibinfo{author}{J.~{Tang}}, \bibinfo{author}{I.~{Kayo}}, and
  \bibinfo{author}{M.~{Takada}}, \bibinfo{journal}{\mnras}
  \bibinfo{volume}{\textbf{416}}, \bibinfo{pages}{2291} (\bibinfo{date}{Sep.
  2011}), \eprint{1103.3614}.
\bibitem{Pueblas:2009}
\bibinfo{author}{S.~{Pueblas}} and \bibinfo{author}{R.~{Scoccimarro}},
  \bibinfo{journal}{\prd} \bibinfo{volume}{\textbf{80}}(4),
  \bibinfo{pages}{043504} (\bibinfo{date}{Aug. 2009}), \eprint{0809.4606}.
\bibitem{Vlah:2012}
\bibinfo{author}{Z.~{Vlah}}, \bibinfo{author}{U.~{Seljak}},
  \bibinfo{author}{P.~{McDonald}}, \bibinfo{author}{T.~{Okumura}}, and
  \bibinfo{author}{T.~{Baldauf}}, \bibinfo{journal}{in preparation} .
\bibitem{Reid:2011}
\bibinfo{author}{B.~A. {Reid}} and \bibinfo{author}{M.~{White}},
  \bibinfo{journal}{\mnras} \bibinfo{volume}{\textbf{417}},
  \bibinfo{pages}{1913} (\bibinfo{date}{Nov. 2011}), \eprint{1105.4165}.
\bibitem{Desjacques:2010a}
\bibinfo{author}{V.~{Desjacques}} and \bibinfo{author}{R.~K. {Sheth}},
  \bibinfo{journal}{\prd} \bibinfo{volume}{\textbf{81}}(2),
  \bibinfo{pages}{023526} (\bibinfo{date}{Jan. 2010}), \eprint{0909.4544}.

\end{thebibliography}
\bibliographystyle{revtex}

\end{document}